\documentclass[journal=jacsat,manuscript=article]{achemso}
\usepackage{chemformula} 
\usepackage[T1]{fontenc}
\usepackage{graphicx}
\usepackage{amsmath}
\newcommand{\angstrom}{\text{\normalfont\AA}}
\usepackage{amssymb}
\usepackage{bm}
\usepackage{caption}
\usepackage{subcaption}
\usepackage{hyperref}
\usepackage{nameref}
\usepackage{booktabs}
\usepackage{multirow}
\usepackage{wrapfig}
\usepackage{comment}

\hypersetup{pdfborder=0 0 0,colorlinks=true,citecolor=black,linkcolor=black}
\input epsf

\newcommand\THEOSMARVEL{Theory and Simulation of Materials (THEOS), and National Centre for Computational Design and Discovery of Novel Materials (MARVEL), {\'E}cole Polytechnique F{\'e}d{\'e}rale de Lausanne, 1015 Lausanne, Switzerland}
\newcommand\UNIBI{Universit{\`a} degli studi di Milano Bicocca, Piazza dell'Ateneo Nuovo 1, 20126 Milano, Italy}
\newcommand\PSI{Laboratory for Materials Simulations, Paul Scherrer Institut (PSI), 5232 Villigen, Switzerland}
\newcommand\BIQUTE{Bicocca Quantum Technologies (BiQuTe) Centre, I-20126 Milano, Italy}

\author{Chiara Cignarella}
\affiliation{\THEOSMARVEL}
\email{chiara.cignarella@epfl.ch}
\author{Davide Campi}
\affiliation{\UNIBI}
\alsoaffiliation{\BIQUTE}
\author{Nicola Marzari}
\affiliation{\THEOSMARVEL}
\alsoaffiliation{\PSI}
\title{Searching for the thinnest metallic wire}

 
\begin{document}

\begin{abstract}

One-dimensional materials have gained much attention in the last decades: from carbon nanotubes to ultrathin nanowires, to few-atom atomic chains, these can all display unique electronic properties and great potential for next-generation applications. 
Exfoliable bulk materials could naturally provide a source for one-dimensional wires with well defined structure and electronics. Here, we explore a database of one-dimensional materials that could be exfoliated from experimentally known three-dimensional Van-der-Waals compounds, searching metallic wires that are resilient to Peierls distortions and could act as vias or interconnects for future downscaled electronic devices. 
As the one-dimensional nature makes these wires particularly susceptible to dynamical instabilities, we carefully characterise vibrational properties to identify stable phases and characterize electronic and dynamical properties.
Our search identifies several novel and stable wires; notably, we identify what could be the thinnest possible exfoliable metallic wire, CuC$_2$, coming a step closer to the ultimate limit in materials downscaling.
\end{abstract}



\clearpage

\section{Introduction}
The rapid development of nanotechnology has led to a growing interest in one-dimensional (1D) materials, due to their unique physical properties and their promise in next-generation applications. For many years, carbon nanotubes had been the poster child of this drive, displaying outstanding properties, but presenting extreme challenges in controlling their chirality and metallic/insulating/semiconducting electronic structure\citep{iijima1991helical,iijima1993single,ajayan1997nanometre,tans1997individual,kim2001thermal}.
Nevertheless, 1D materials remain extremely compelling in providing an effective route towards ultimate downscaled electronic components \cite{geremew2018current,stolyarov2016breakdown} and outstanding candidates for flexible electronics\cite{Park2020}. Moreover, fundamentally fascinating phenomena such as the emergence of charge-density waves\citep{gruner2018density,peierls1955quantum}, Luttinger liquid\citep{luttinger1963exactly,haldane1981luttinger,voit1995one}, or exciton condensation\citep{Varsano2017} can arise from dimensionality reduction.

Different techniques of growing 1D nanowires or even single-atom chains on suitable substrates have been developed in recent years, through either self-assembly or direct growth\citep{gambardella2002ferromagnetism,crain2004chains,zeng2008charge,ferstl2016self,qin2018epitaxial,guo2022direct}, with the caveat that non-negligible interactions with the surface can modify the purely 1D features and change potential applications\citep{sanna2018one,yogi2022electronic}. As mentioned, carbon nanotubes\citep{iijima1991helical,iijima1993single,ajayan1997nanometre,tans1997individual,kim2001thermal} (CNTs) represented the most remarkable early example of one-dimensional systems, but the difficulty in controlling chirality (and therefore electronic properties) during CNTs growth has so far prevented their widespread use in nanotechnology.
In addition to CNTs, in the last few years research in one-dimensional systems has made significant advances: more complex methods have succeeded to obtain ultrathin nanowires, from the encapsulation of atomic chains in carbon nanotubes\citep{stonemeyer2020stabilization,kashtiban2021linear} to directing-agents synthesis\citep{yan2017hybrid, xiao2018electrically} that make it possible to achieve wires down to only a 3-atom diameter, delivering the smallest freestanding inorganic nanowires so far\citep{yan2017hybrid}. 

In this pursuit, three-dimensional crystals in which inorganic wires are held together by weak Van-der-Waals interactions\citep{balandin2022one,meng2022one} could be very promising. Similarly to what is done for two-dimensional monolayers\citep{novoselov2012roadmap,wang2012electronics, nicolosi2013liquid,hanlon2015liquid}, one-dimensional individual wires could be potentially isolated from the bulk by mechanical or chemical exfoliation \citep{stolyarov2016breakdown,island2017electronics,lipatov2018quasi,barani2021electrically,kargar2022elemental} in order to obtain wires of few-atom width.
Stable atomic wires would intrinsically lack surface dangling bonds and edge scattering, making them suitable for ultimate-scaled electronics\citep{meng2022one,balandin2022one}; most importantly, and at variance with the chirality dependence of CNTs, they would display well defined and reproducible electronic properties.

Uncovering novel one-dimensional systems is therefore of great interest. High-throughput (HT) studies already provided a rich portfolio of promising and novel 2D materials that are exfoliable into two-dimensional layers\citep{lebegue2013two, ashton2017topology, choudhary2017high, cheon2017data, mounet2018two, haastrup2018computational, larsen2019definition, campi2023expansion}; using similar approaches to discover new atomic wires is also an emerging new field\citep{shang2020atomic,zhu2021spectrum, moustafa2022computational,moustafa2023hundreds}.
Within this framework, we performed a high-throughput screening of experimentally known compounds to identify a database of more than 800 1D systems that could be exfoliated from experimentally known three-dimensional bulks (typically Van-der-Waals). Here, we focus our search on metallic 1D wires; these could be precious for applications such as vias or interconnects in field-effect transistors (FET), with the transition-metal trichalcogenide TaSe$_3$ being an outstanding example\citep{stolyarov2016breakdown,liu2017low,empante2019low}, or for ultrashort gate electrodes for ultimate FET-length scaling\citep{desai2016mos2}.  
Metallic one-dimensional systems represent moreover a powerful platform to study real-world one-dimensional physics, such as Peierls transitions and charge-density waves beyond ideal systems or models.

Carbyne, the infinite linear carbon chain, has been studied since long time as the ultimate metallic one-dimensional system. In its symmetric phase of identical carbon-carbon bonds, called \textit{cumulene}, carbyne is metallic; however, due to a Peierls distortion it dimerises into the \textit{polyyne} phase, with bond-length alternation and the opening of a gap. Even if predicted to have exceptional properties, its realization is extremely challenging and the existence of carbyne itself has long been debated. The first synthesis of a finite polyyne chain 44-atom long was achieved in 2010\citep{chalifoux2010synthesis} and, successively, different methods have been investigated to synthesise short chains capped with chemical groups, or longer chain enclosed in double or multi-walled CNTs to overcome stabilization issues\citep{shi2016confined,buntov2019structure}. Experimental investigations have remained very limited, due to high chemical reactivity with respect to the environment and poor stability in ambient conditions, that make carbon chains difficult to produce and 
characterise. In addition, electronic properties seems very sensitive to the diameter of the confining CNTs\citep{shi2016confined} and the metallic contacts for measurements\citep{la2015strain}, making production with well-defined properties difficult.
Obtaining metallic cumulene is even more challenging; albeit recent theoretical work\citep{artyukhov2014mechanically} shows that anharmonic quantum vibrations help to stabilise it against Peierls distortions, this metallic phase is still metastable\citep{romanin2021dominant}, while polyyne remains the most stable phase even at very high temperatures.

Exfoliable wires from Van-der-Waals bulk crystals can therefore provide a promising route in finding new stable one-dimensional systems.
Here, we characterise the dynamical stability of the most promising systems in our database, analysing their phonon dispersions, and, when possible, finding the appropriate superstructure for which the wire is stable.
Electronic and mechanical properties are then investigated in view of possible applications in nanotechnology, together with a more detailed analysis for all stable materials preserving a metallic phase.
In search for the building blocks for downscaled electronic devices we find novel exfoliable atomic wires that are mechanically stable against these instabilities. Remarkably, we identify the metallic CuC$_2$ in addition to TaSe$_3$ and two novel semi-metallic Sb$_2$Te$_2$ and Ag$_2$Se$_2$.

\section{Results and discussions}
\subsection{Selection of the candidates}
This work starts from a large database (Campi D.; Marzari N., \textit{in preparation}) of more than 800 1D materials that could be exfoliated from experimentally known, Van-der-Waals bonded, bulk materials; these are sourced from three different databases: the Crystallographic Open Database (COD)\citep{COD}, the Inorganic Crystal Structure Database\citep{ICSD,bergerhoff1983inorganic} and the Pauling File\citep{mpds,villars1998linus}. The identification of these materials follows a protocol adapted from previous studies on 2D materials\citep{mounet2018two,campi2023expansion}: here, for all the structures with up to 20 atoms per unit cell and up to 4 different atomic species, we first identify chemically connected sub-structures using a very tolerant, purely geometrical, criterion comparing interatomic distances of each atom pair with the differences between their respective van-der-Waals radii, and considering such pairs chemically connected if their distances are lower than a certain range of thresholds identified in Ref.\citenum{alvarez}. The N-th dimensionality of the sub-structure is then inferred from the rank of the matrix formed by all the vectors connecting an atom with all its chemically connected replica. For these candidates binding energies are then computed from first-principles as the energy difference per unit length between the bulk structure optimized at its theoretical equilibrium parameter using a van der Waals functional (vdw-df2-c09)\citep{lee2010,cooper2010,hamada2010}, and the energy of each isolated substructure. The geometry of the 1D wires is then optimized for the isolated systems, and the electronic and dynamical properties are computed with the PBE functional\citep{PBE}. 
Here, we screen this portfolio of 1D materials for metallic wires; from those, we select only those appearing dynamically stable from the phonon dispersions computed within the aforementioned high-throughput screening. 

In the screening we decided not to use binding energies a priori as an exclusion criteria, meaning that we included a material regardless of its computed binding energy; even though it is intuitive to imagine a correlation between the binding energy and ease to isolate the material, it is difficult to draw a reliable line between a material that could or could not be experimentally isolated based solely on its binding energy, also due to the lack of experimental data.
Last, we add TaSe$_3$ to the materials selected following the aforementioned procedure because of its successful experimental exfoliation in very thin bundles \citep{stolyarov2016breakdown,empante2019low,liu2017low} that makes it an interesting benchmark for our study, albeit our screening classifies it initially as a 2D sheet rather than a 1D wire. TaSe$_3$ in a quasi-1D form  has also proven to be a remarkable good candidate for local interconnects, with a breakdown current up to 10$^8$ A/cm$^{2}$\citep{empante2019low}.

\begin{figure}[h!]
    \centering
    \hspace*{-0.5cm} 
        \includegraphics[scale=0.12]{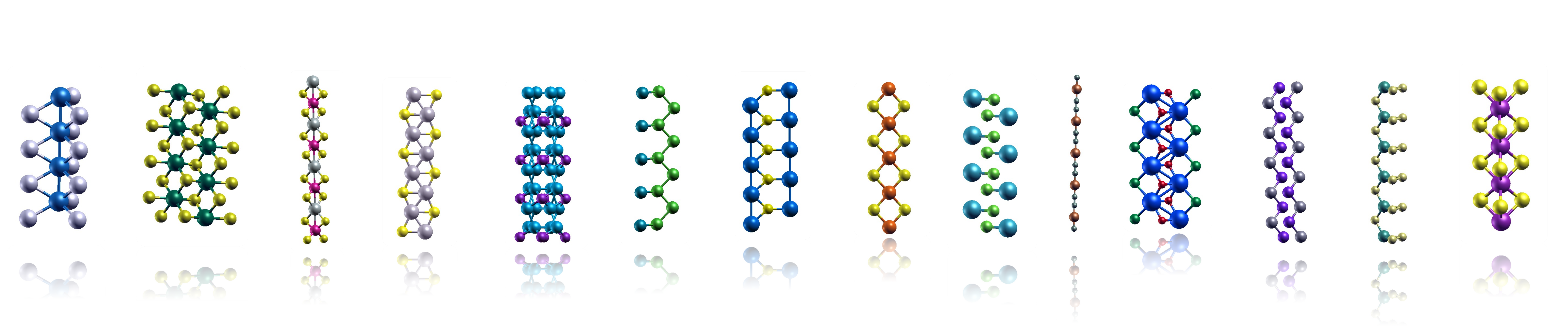}
    \caption{The 14 1D wires considered in the present study, as exfoliable from experimentally know 3D inorganic materials. From left to right (exfoliation energy for reference, in meV/\AA ): CdI$_3$ (1203), Hf$_2$Se$_6$ (4287), AgS$_4$W (880), Ag$_2$Se$_2$ (1735), As$_4$Pd$_8$ (5399), ITe$_2$ (891), Pb$_2$Se (6544), AuSe$_2$ (588), Bi$_2$S$_2$ (1787), CuC$_2$ (636), Cl$_2$O$_2$Pb$_2$ (579), Sb$_2$Te$_2$ (1560), SnS$_3$ (2427), TaSe$_3$ (our benchmark; see text). }
    \label{chains}
\end{figure}

The portfolio of 14 candidate structures is presented in Figure \ref{chains}.
For each system we consider at first the structure at equilibrium in the primitive unit cell as isolated from the three-dimensional parent, with the inherited lattice periodicity and symmetries, followed by geometrical relaxation and calculation of the phonon dispersions and eventual instabilities.
We note that structural optimizations and phonon calculations in 1D metallic systems are extremely sensitive to numerical details: they require very dense \textbf{k}- and \textbf{q}- points sampling in the Brillouin zone for the electronic and vibrational states to accurately describe the electronic states at the Fermi level that might drive a dynamical instability. Also, an adequate amount of vacuum across the two directions perpendicular to the wire ($\sim$25 \AA) is needed to avoid long-range spurious interactions\citep{kozinsky2006static} among the periodic replicas. 
However, high-throughput screenings performed on thousands of materials necessarily need a trade-off between accuracy and computational cost. Thus, with the use of relatively coarse grids for Brillouin zone sampling, some materials, and in particular metallic systems, could appear stable and an eventual instability can be revealed only with the use of more refined parameters. 
For the materials selected, we therefore recomputed all the properties employing much stricter parameters with respect to the initial high-throughput screening to simulate our systems accurately (up to 1x1x300 k-point sampling and up to 27~\angstrom~of vacuum; these are as described in detail in the \nameref{methods} section).

\subsection{Dynamical stability analysis}\label{dynamical_analysis}
One-dimensional metallic wires are known to be particularly susceptible towards electronic and dynamical instabilities. 
In his pioneering work in 1930, Peierls\citep{peierls1930theorie,peierls1955quantum} argued that an ideal half-filled one-dimensional chain of atoms is always unstable to any periodic potential with a wavevector 2k$_F$ two times the Fermi wavevector, and spontaneously distorts into a dimerised chain, opening a gap at the zone boundaries of the new unit cell. 
This happens since the electron energy gain ($\sim u^2$ ln$u$) exceeds the cost of the atomic distortion ($\sim u^2$) so that the transition is energetically favourable below a certain critical temperature \citep{pouget2016peierls, khomskii2010basic}.
The atomic distortion comes together with the modulation of the electronic charge density with the same wavevector 2k$_F$, $\langle\rho(x)\rangle = \rho_0 + \rho_1 e^{i2k_Fx}$, \textit{i.e.}, the formation of a charge-density wave (CDW)\citep{gruner1988thedyn,frohlich1954theory}. For instance, in an ideal dimerised chain the electron density oscillates between two consecutive bonds in a doubled unit cell.
Few years later Kohn\citep{kohn1959image} pointed out that this instability is reflected in the phonon frequencies which become imaginary as soon as one approaches q=2k$_F$ below the critical temperature; this dip in the phonon spectrum is know as a Kohn anomaly (KA) \citep{zhu2015classification,zhu2017misconceptions}.
In an ideal one-dimensional scenario, the static response function $\chi_0$ displays a logarithmic divergence at wavevector 2k$_F$ $\chi_0(\mathbf{q},\omega=0)\sim$ ln$|$q-2k$_F|$, where the chain has the nesting of the Fermi surface (FSN). Such divergence implies that the system experiences a potential instability when exposed to any external perturbation at the corresponding wavevector \citep{toombs1978quasi}. However, in more realistic 1D cases, such FS nesting can appear at any wavevector q, with consequent Kohn anomaly.
Realistic freestanding 1D systems can show a variety of behaviours deviating from the classical Peierls mechanism of the ideal atomic chain\citep{zhu2017misconceptions}.
For example, the overlap term $|\langle \psi_{\mathbf{k}} | e^{i\mathbf{q}\cdot\mathbf{r}} | \psi_{\mathbf{k+q}}\rangle|^2$, which is typically neglected in the treatment of Peierls instabilities in ideal systems \citep{johannes2006fermi}, proves to be important in cases with a mixed-character band crossing around the Fermi energy, as for the linear Li chain \citep{Derriche2022Suppression} where this term can smooth the response and suppress the classical nesting mechanism of instability, thereby stabilizing the system.
Moreover, it is plausible that the modulations in the electron-phonon coupling strength might play a non-negligible role, as it has been observed in 2D and 3D systems  \citep{johannes2006fermi,johannes2008fermi,calandra2009effect,diego2021van,zhu2015classification}.

Here, in order to reveal any potential structural instability, signaled by a negative phonon mode at some \textbf{q}-point of the Brillouin zone, we systematically compute phonon dispersions in the initial configuration by means of density-functional perturbation theory (DFPT)\citep{baroni2001phonons} over very fine grids.  
Most of the systems, but notably not all, do show negative (imaginary) phonon frequencies in the initial structure where the primitive unit cell is inherited from the 3D parent once phonons are computed with sufficient accuracy (all the phonon dispersions computed in this work are reported in the Supporting Information).
This means that these would not be stable as isolated freestanding wires, in the configuration extracted from the bulk, and would prefer to dimerise or more generally relax or reconstruct in a supercell structure with a longer periodicity; in the most extreme cases, these instabilities could also be a sign that the system cannot exist in 1D form.

\begin{figure}[h!]
    \centering
    \begin{subfigure}{0.5\textwidth}
    \centering
    \includegraphics[scale=0.53]{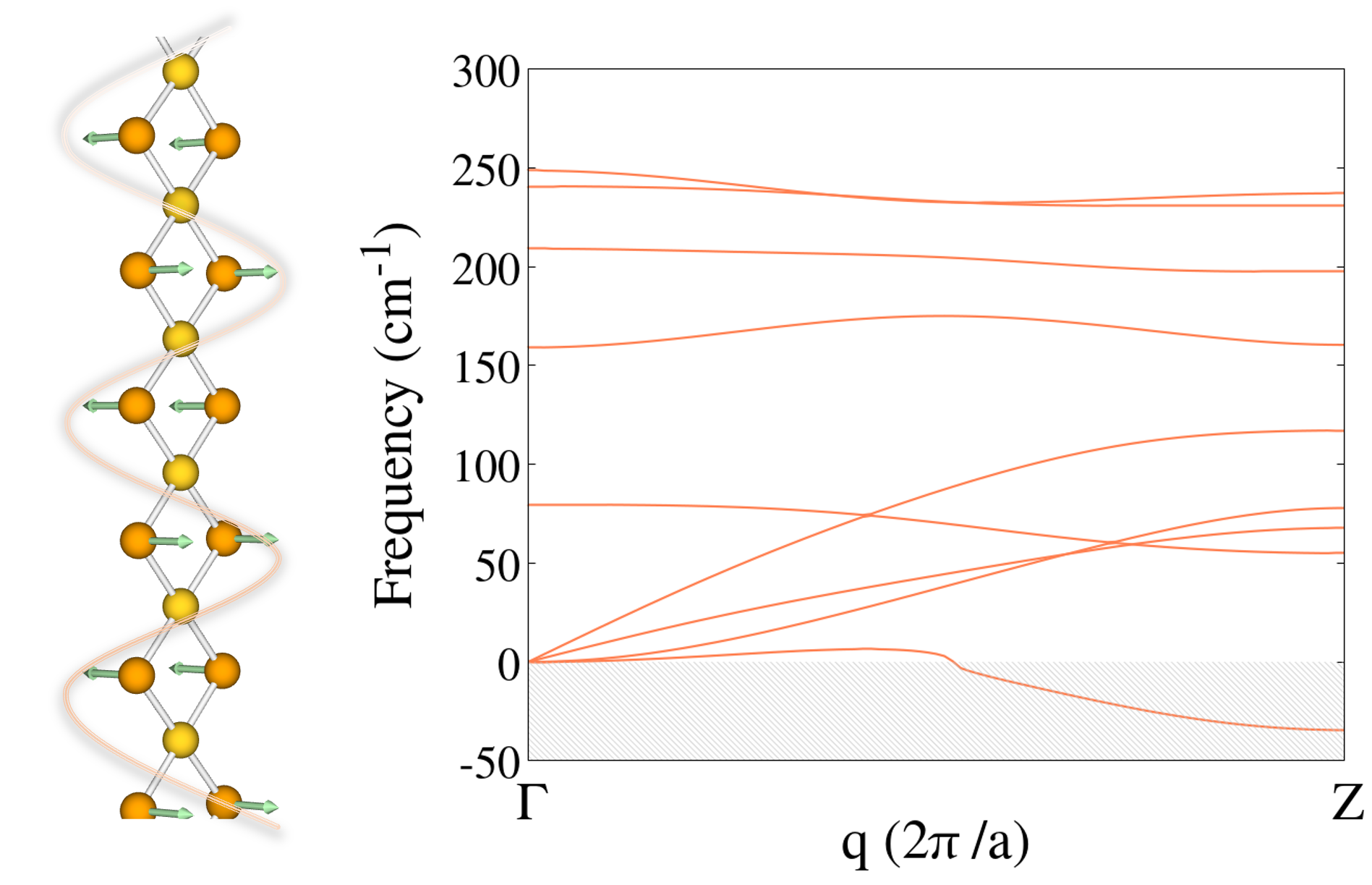}
    \caption{}
    \end{subfigure}
\hspace{5mm}
    \begin{subfigure}{0.4\textwidth}
    \centering
    \includegraphics[scale=0.53]{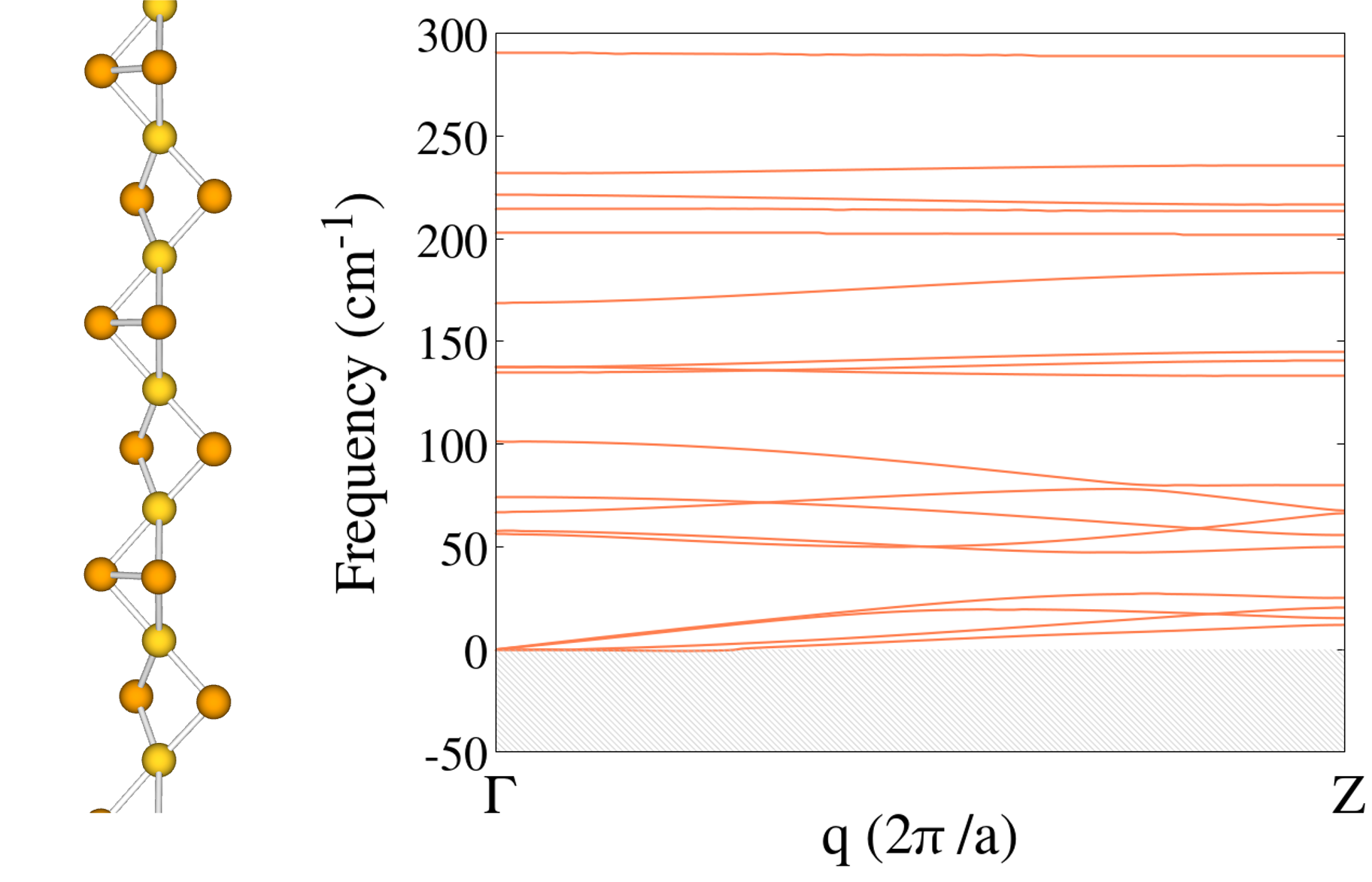}
    \caption{}
    \end{subfigure}%
    \caption{Phonon dispersion for AuSe$_2$ (Au: yellow, Se: orange) (a) in the configuration isolated from the 3D bulk; the green arrows indicate the eigenmodes of the unstable phonon in Z.
    (b) in the final relaxed and reconstructed structure, in a doubled supercell.}
    \label{au}
\end{figure}

In order to find, for each material, the most stable structure, we relax the atoms in a supercell commensurate with the wavevectors of the lowest negative phonon. In addition, to break the original symmetry, atoms are initially displaced following the eigenvectors of the corresponding soft phonon  $\mathbf{q}$ (see \nameref{methods}).
For example, in their "from-the-bulk" primitive cell, AuSe$_2$ and Ag$_2$Se$_2$ display a broad phonon softening with minimum at a $Z$=(0,0,1/2) [in units of $2\pi/a$]; this might disappear in a 1 $\times$ 1 $\times$ 2 supercell, \textit{i.e.}, allowing for the doubling of the unit cell along the direction of the chain and subsequent relaxation in these two cases. The unstable modes mainly come from the out-of-plane (\emph{xy} plane) vibrations of selenium atoms (see Fig. \ref{au}) resulting in a "zig-zag" reconstructed and dynamically stable structure that can be obtained after relaxing the system in a doubled supercell.
Other compounds like AgWS$_4$, As$_4$Pd$_8$ and Bi$_2$S$_2$ soften at different $\mathbf{q}$-points along the $\Gamma$-$Z$ path, requiring larger superstructures to account for relaxation and sometimes reconstruction; in some cases, at wavevectors incommensurate with the lattice.
Two systems, Pb$_2$Se and CdI$_3$, remain unstable even after multiple relaxations along the negative phonon modes, and for this reasons we exclude them as possible target compounds.
The results of this analysis are summarized in Table \ref{tab:summary}, together with the 1 $\times$ 1 $\times$ \emph{n} supercell used for the relaxation.
We observe that most of the materials became insulating after reconstruction; in this case we report the energy gap E$_g$ that has opened. Most notably, however, our search finds also stable metallic materials that remain in the configuration inherited from the three-dimensional crystal: namely, CuC$_2$, TaSe$_3$, and Sb$_2$Te$_2$, and Ag$_2$Se$_2$, that is unstable in its initial structure but remains metallic after reconstruction. These four materials are therefore extremely appealing as possible novel ultrathin conducting nanowires, targeting highly efficient downscaled electronic devices and flexible electronics. 
Moreover, from the fundamental point of view, one-dimensional metallic systems could exhibit Luttinger liquid behaviour, characterised by collective motions of electrons and charge-spin separation\citep{luttinger1963exactly,haldane1981luttinger,voit1995one}; experimental findings of this regime in real materials are still scarce\citep{bockrath1999luttinger, Chang2003chiral, blumenstein2011atomically} and the stable metallic 1D materials identified in this study could represent an additional experimental platform. 

It is worth noting that the present stability analysis does not consider quantum anharmonicity, which can be helpful in certain freestanding cases. For example, in highly anharmonic systems or in case of light atoms like carbyne, zero-point motion reduces or suppresses the formation of  CDW\citep{romanin2021dominant,artyukhov2014mechanically,bianco2019quantum}. While exploring this aspect in detail would be intriguing, and could result in additional stable 1D metallic wires, it goes beyond the scope of the current study.
Finally, our study refers to freestanding wires at zero T; however, identifying the critical temperature of the insulator-to-metal transition\citep{bianco2020weak,diego2021van,bianco2019quantum} and/or the effect of the substrate\citep{sanna2018one,bianco2019quantum} could also be relevant for practical applications.

\subsection{Electronic and mechanical properties}\label{mec_and_ele}
In order to analyse the electronic properties of the compounds, we compute the band structure and the density-of-states for each material in its original and reconstructed form, and summarise the results in Table \ref{tab:summary}. The coordinates of the original cell and the reconstructed cell for each of the system are reported in the Supporting Information, as well as the electronic band structures. 

\begin{table}[h!]
\centering
    \begin{tabular}{l c c c c r}
          & DOS$\big\lvert_{E_F}$ (meV \AA$^3$)$^{-1}$ & &\multicolumn{2}{c}{DOS$\big\lvert_{E_F/HOMO}$(meV \AA$^3$)$^{-1}$}&E$_g$ (eV) \\
\midrule
          &\emph{initial structure}&supercell size&\\
          &\emph{(from 3D bulk)}&&\\
\midrule
        CuC$_2$ &41&1$\times$ 1 $\times$ 1&41 &(E$_F$)& metal\\
        TaSe$_3$ &14&1$\times$ 1 $\times$ 1&14&&metal\\
        Sb$_2$Te$_2$& Dirac cone&1$\times$ 1 $\times$ 1 &Dirac cone&&semi-metal \\
        Ag$_2$Se$_2$&400&1 $\times$ 1 $\times$ 2& Dirac cone&&semi-metal* \\
\cmidrule(lr){1-6}
        Cl$_2$O$_2$Pb$_2$ &855&1 $\times$ 1 $\times$ 2&1798& (HOMO)&1.80\\
        AuSe$_2$ &61&1 $\times$ 1 $\times$ 2&1400 &&0.48\\
        AgWS$_4$ &37&1 $\times$ 1 $\times$ 6&1030&&0.25\\
        As$_4$Pd$_8$&17&1 $\times$ 1 $\times$ 4&642&&0.11\\
        ITe$_2$ &26&1 $\times$ 1 $\times$ 2 &114&&0.74\\
        Hf$_2$Se$_6$ &10&1 $\times$ 1 $\times$ 1&47&&0.67\\
\cmidrule(lr){1-6}
         Bi$_2$S$_2$ & 13 &1 $\times$ 1 $\times$ 5&\multicolumn{3}{r}{dissociates}\\
         SnS$_3$ &Dirac cone&1 $\times$ 1 $\times$ 5&\multicolumn{3}{r}{dissociates}\\
       Pb$_2$Se & 10&1 $\times$ 1 $\times$ 2&\multicolumn{3}{r}{unstable phonons}\\
       CdI$_3$ & 9&1 $\times$ 1 $\times$ 3&\multicolumn{3}{r}{unstable phonons}\\
\bottomrule
    \end{tabular}
    \caption{Stability analysis of the 14 one-dimensional materials studied, with the stable supercell 1 $\times$ 1 $\times$ \emph{n} reported, followed by the density-of-states per unit volume in the initial and relaxed phases, reported at the Fermi level for metallic materials and at the HOMO for insulators, and band gaps for the latter computed at the Kohn-Sham PBE level. The unit volume has been calculated (see text) using the quantum volume definition of Ref. \citenum{cococcioni2005electronic}. (*Ag$_2$Se$_2$ opens a small gap with SOC, as shown in next section.)}\label{tab:summary}
\end{table}

Besides the two metallic materials and the two semi-metals, we have six systems stable in the insulating phase, for which we show the density of states DOS$\big|_{HOMO}$ in the final relaxed configuration.
We observe overall an increase of this DOS$\big|_{HOMO}$ from the initial to the reconstructed phase, mainly due to the presence of very flat bands in the insulating reconstructed structures (electronic band structures are reported in the Supporting Information). Albeit in the present study we focus on metallic one-dimensional systems, semiconductor wires are equally interesting from an application point of view. In addition, the materials in our study that undergo a metal-insulator transitions upon reconstruction could also be useful in electronic devices\citep{liu2016charge,zhu2018light,geremew2019bias,mohammadzadeh2021evidence} where the transition could be tuned by different effects such as temperature, electric fields or mechanical distortions.

In the light of possible applications for flexible electronics\citep{gao2022strongest,treacy1996exceptionally}
we further characterise the systems computing their mechanical properties. 
We can assess the tensile stiffness through the Young modulus, \textit{i.e.}, the response of the material to stresses applied along one axis; in this case, the periodic direction of the chain. In Table \ref{young} the Young moduli calculated are reported, where we calculated the volume of the wires as the quantum volume defined in Ref.\citenum{cococcioni2005electronic} (details on the computation are in section \nameref{methods}). 
\begin{table}[h!]
\centering
    \begin{tabular}{c l }
          &Y (GPa)\\
\toprule
        CuC$_2$ &242\\
        TaSe$_3$ & 75\\
        Sb$_2$Te$_2$ &20\\
        Ag$_2$Se$_2$ &23 (41) \\
        \cmidrule(lr){1-2}
        Cl$_2$O$_2$Pb$_2$  &8 (26)\\
        AuSe$_2$ & 11 (52)\\
        AgWS$_4$ & 36 (38)\\
        As$_4$Pd$_8$& 23 (28)\\
        ITe$_2$ &14 (8)\\
        Hf$_2$Se$_6$ &48 (30)\\
\bottomrule
    \end{tabular}
    \caption{Young moduli Y for the stable phases of the 10 stable materials. In parenthesis, $Y$ for the unstable metallic phase at 0 K.}\label{young}
\end{table}
The Young moduli range obtained, between approximately 10 and 80 GPa, aligns with the values reported for inorganic nanowires\citep{leu2008ab} and the single wire of TaSe$_3$, where the Young modulus is analogous to the three-dimensional anharmonic crystal\citep{tritt1994measure}. Notably, CuC$_2$ exhibits a remarkable Young modulus of 242 GPa, becoming comparable to those of CNTs \citep{treacy1996exceptionally}. This range of Young moduli holds promise for the development of conductive materials suitable for flexible electronics and applications involving strain engineering, where one-dimensional materials show numerous advantages\citep{liu2015flexible,Park2020}. 

As a final analysis, we inspect possible presence of magnetism for the stable metallic and semi-metallic systems. Theoretical work found indeed magnetic ground-state with half-metallic properties in finite linear atomic chains of carbon-transitional-metal compounds\citep{dag2005half}.
We relax the systems with starting ferromagnetic or antiferromagnetic initial configurations, with both spins parallel and transversal to the direction of the wire; we didn't observe magnetism in any of the wires (details in the Method section). 

\subsection{Stable metallic materials at 0K}\label{details_stable}
In the following, we explore in greater detail the metallic chains that remain stable at 0 K, where we have identified two particularly fascinating cases: CuC$_2$ and Ag$_2$Se$_2$. CuC$_2$ provides the thinnest stable metallic wire that can be isolated from experimentally-known 3D bulk systems, while Ag$_2$Se$_2$ shows promising indications of being a topological one-dimensional material.
\subsubsection{CuC$_2$ as the thinnest metallic wire}
CuC$_2$ is a straight-line chain composed by two carbons and one copper, and, with a diameter computed by means of the quantum volume of only \emph{d}=4 \AA, represents the thinnest freestanding metallic nanowire stable at 0 K. 

\begin{figure}[h!]
    \centering
    \includegraphics[scale=0.70]{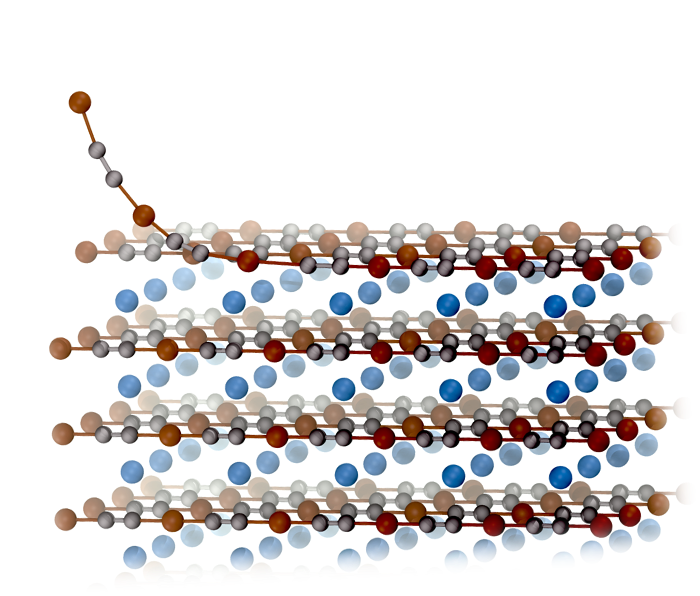}
    \caption{One of the three-dimensional parent of CuC$_2$, the exfoliable alkali metal carbide NaCuC$_2$\citep{ruschewitz2003binary} (Na: blue, Cu: brown; C: grey), with the CuC$_2$ chains running parallel. Other parents include KCuC$_2$ and RbCuC$_2$\citep{ruschewitz2003binary,cremer2002alkali}.}
    \label{nacuc2}
\end{figure}

CuC$_2$ was recently synthetised\citep{sun2016bottom} as result of metalation of carbon chains from ethyne precursors on the Cu(110) surface in ultrahigh vacuum conditions and the finite metalated chain was proposed as a promising molecular wire\citep{tu2016cu}.
Notably, here we found that the material is exfoliable starting from multiple three-dimensional Van der Waals crystals, making the production potentially simpler. In particular, it can be extracted from three compounds of the family of experimentally known ternary alkali-metal carbides:  NaCuC$_2$, KCuC$_2$ and RbCuC$_2$\citep{ruschewitz2003binary,cremer2002alkali} (Fig.\ref{nacuc2}); our calculations show that it can be exfoliated from KCuC$_2$ with a moderate binding energy of 636 meV/\AA, comparable with that of known quasi-1D materials like $\alpha$-tellurium (376 meV/\AA) and Sb$_2$S$_3$ (528 meV/\AA).

\begin{figure}[h!]
    \centering
    \begin{subfigure}{1\textwidth}
    \centering
    \includegraphics[scale=0.21]{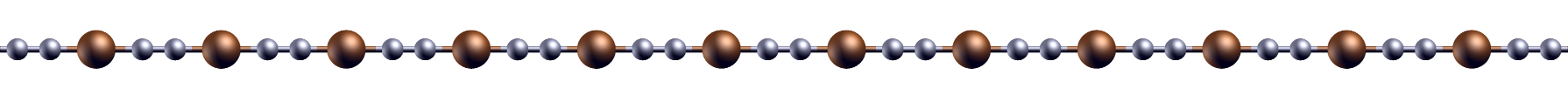}
    \end{subfigure}
    \vspace{0.3cm}
    \begin{subfigure}{0.45\textwidth}
    \centering
    \includegraphics[scale=0.21]{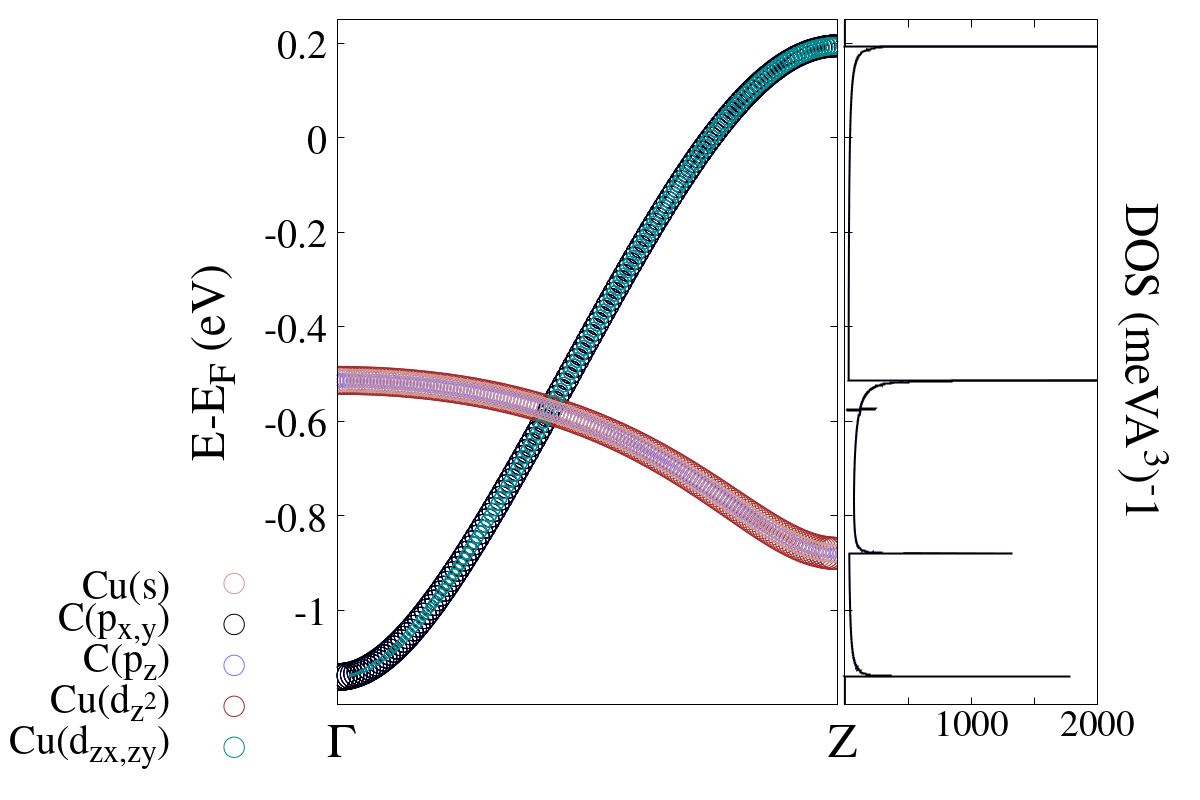}
    \caption{}
    \end{subfigure} \hspace{1cm}
    \begin{subfigure}{0.45\textwidth}
    \includegraphics[scale=0.20]{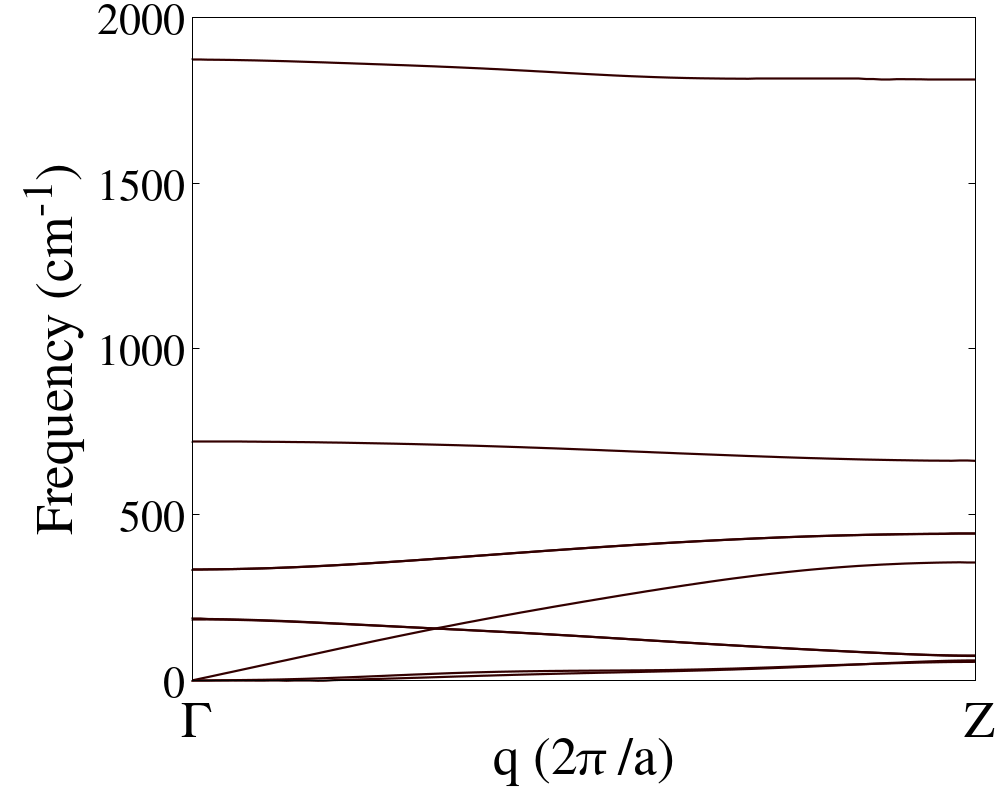}
    \caption{}
    \end{subfigure}
    \caption{One-dimensional CuC$_2$ (Cu: brown; C: grey): (a) electronic band structure with projected density of states. At the Fermi surface the most contributions comes from C-\emph{p$_x$}, C-\emph{p$_y$} and to a lesser extent Cu-\emph{d$_{zx}$} and Cu-\emph{d$_{zy}$}. (In the plots we show only the most relevant orbital contributions.) (b) Stable phonon dispersions along the Brillouin zone.}
    \label{c2cu}
\end{figure}

\begin{figure}[h!]
    \centering
    \includegraphics[scale=0.50]{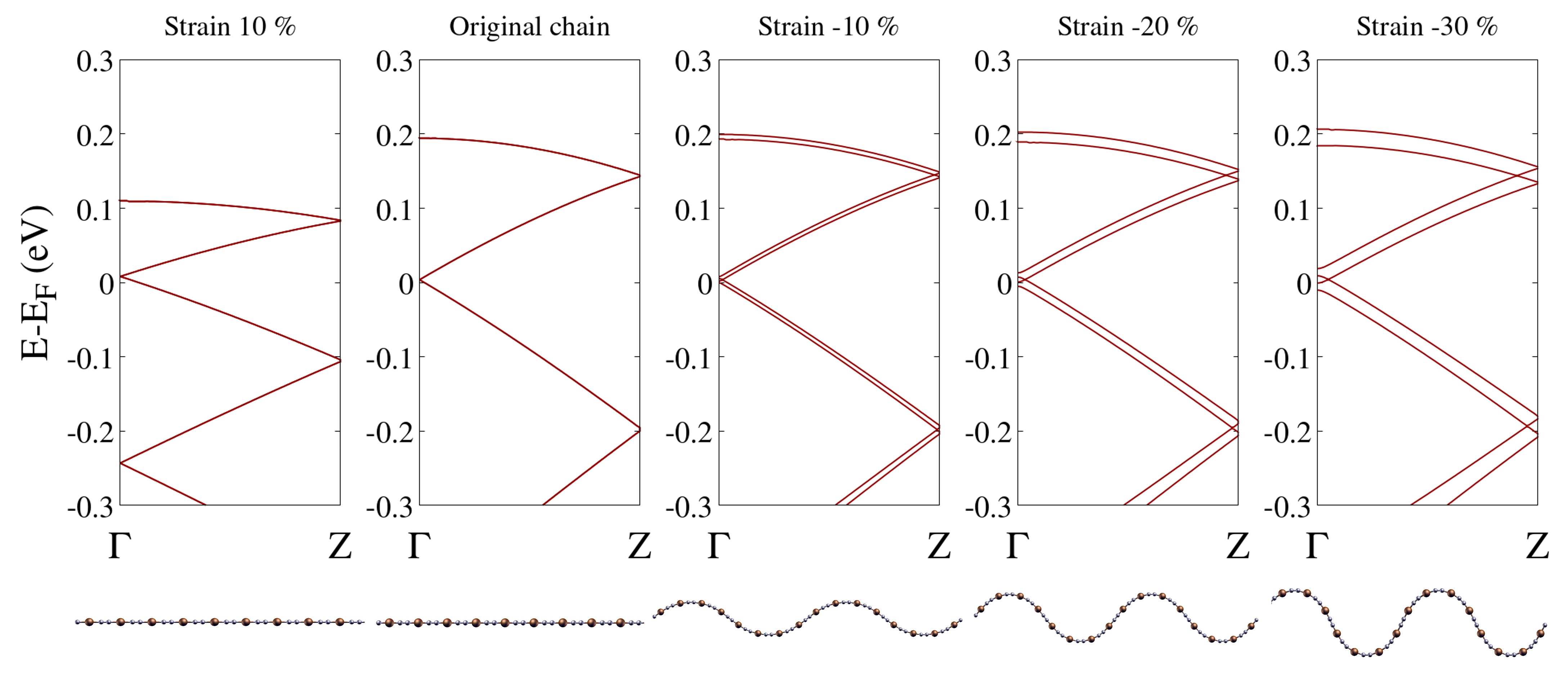}
    \caption{Bent CuC$_2$: electronic band structures for different strains, and final relaxed structures. Positive strain corresponds to elongation, negative strain to compression; the system remains metallic in all the cases studied.}\label{strain_bend_c2cu}
\end{figure}

The single wire of CuC$_2$ is dynamically stable in its phase isolated from the bulk, as seen in the phonon dispersions along the Brillouin zone (Fig. \ref{c2cu} (b)).  
In Figure \ref{c2cu} (a) we report the electronic band structure and density of states; we observe the typical shape of the DOS for 1D materials, with peaks corresponding to the Van Hove singularities (VHSs). 
CuC$_2$ has the highest DOS$\big|_{E_F}$ within the materials studied (41 /meV \AA$^3$), as well as the largest Young modulus (242 GPa). 
With an eye to possible applications in flexible electronics, we investigate the behavior of the system under bending; we first apply a compressive/extensive strain to the original wire, while subjecting it to a sinusoidal bending within an 8-cell supercell (details in the \nameref{methods} section).
As expected, when the chain is stretched (corresponding to the positive value of strain in Fig. \ref{strain_bend_c2cu}), the system recovers its linear shape. When compressed, it relaxes maintaining the sinusoidal modulation while remaining flat in the perpendicular x-y plane; most notably, in all cases the metallic character is fully preserved.
The results are illustrated in Fig. \ref{strain_bend_c2cu}: we observe that CuC$_2$ remains metallic even under significant bending, making it particularly appealing for the applications (when bending, the electronic bands split into two, which can be attributed to higher and lower degree of strain between the curved and linear segments). As it can be seen, even extreme curvatures/bending are possible, still preserving metallicity.
In the Supporting Information (SI) we provide additional details, including band structures under different strains (without bending) observing again that metallicity is preserved. In particular, we study the reactivity against molecular oxygen O$_2$, where we notice at most a weak physisorption on the chain and no chain breakage or chemisorption of O$_2$. 
We further verify the electronic structure of the CuC$_2$ wire using hybrid-functional calculation; a comparison of the band structures calculated with PBE and with HSE06 is presented in the SI. The overall effect of the hybrid functional is a downward shift of the valence bands but the band crossing at the Fermi energy persists, thereby preserving the metallic character, as also expected by electron counting (27 $e^{-}$).

Experimental exfoliation of this exceptional material would therefore bring us closer to realising the thinnest metallic wire possible.
 
\subsubsection{Other metal wires: Ag$_2$Se$_2$, Sb$_2$Te$_2$ and TaSe$_3$}
\begin{figure}[h!]
    \centering
    \begin{subfigure}{1\textwidth}
    \centering
    \includegraphics[scale=0.20]{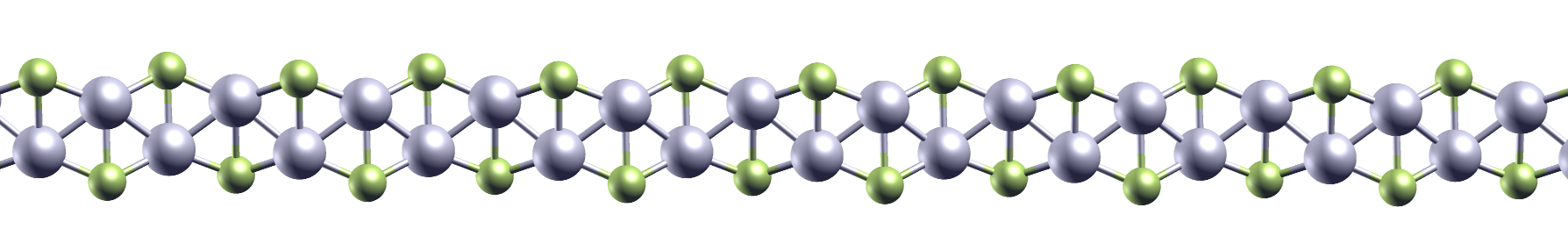}
    \end{subfigure}
    \vspace{0.3cm}
    \begin{subfigure}{0.45\textwidth}
    \centering
    \includegraphics[scale=0.21]{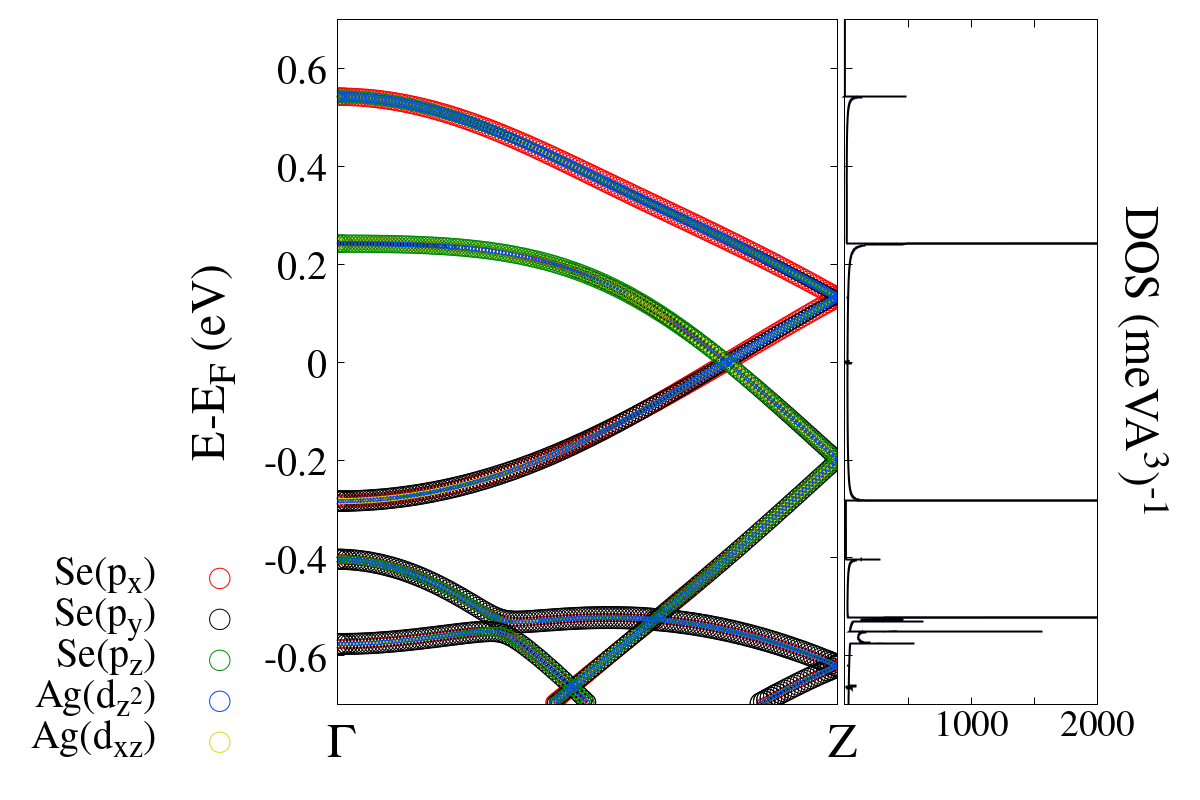}
    \caption{}
    \end{subfigure} \hspace{1cm}
    \begin{subfigure}{0.45\textwidth}
    \includegraphics[scale=0.20]{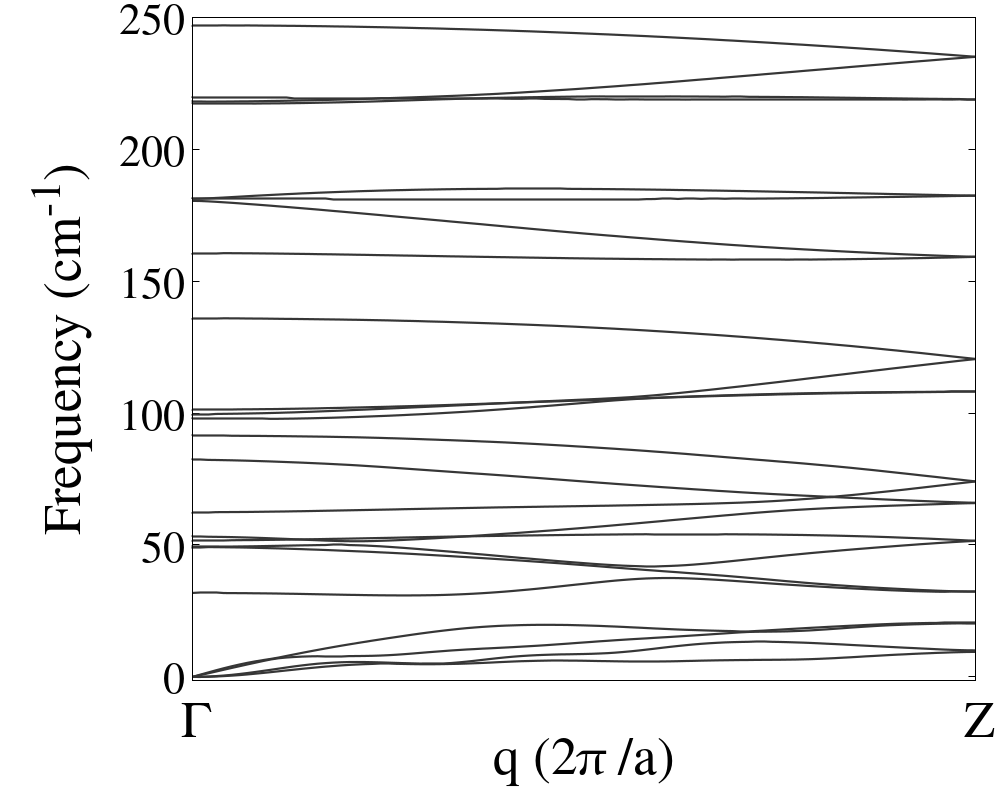}
    \caption{}
    \end{subfigure}
    
    \hspace{1cm}
    \begin{subfigure}{0.45\textwidth}
    \centering
    \includegraphics[scale=0.20]{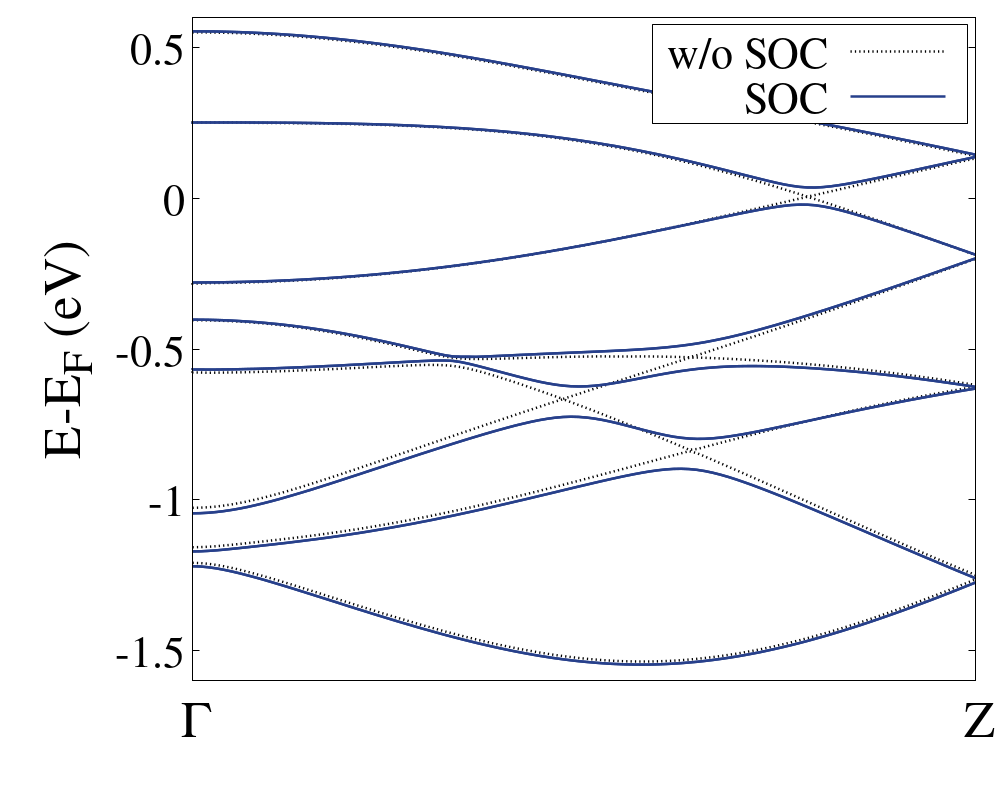}
    \caption{}
    \end{subfigure}
    \begin{subfigure}{0.45\textwidth}
    \includegraphics[scale=0.20]{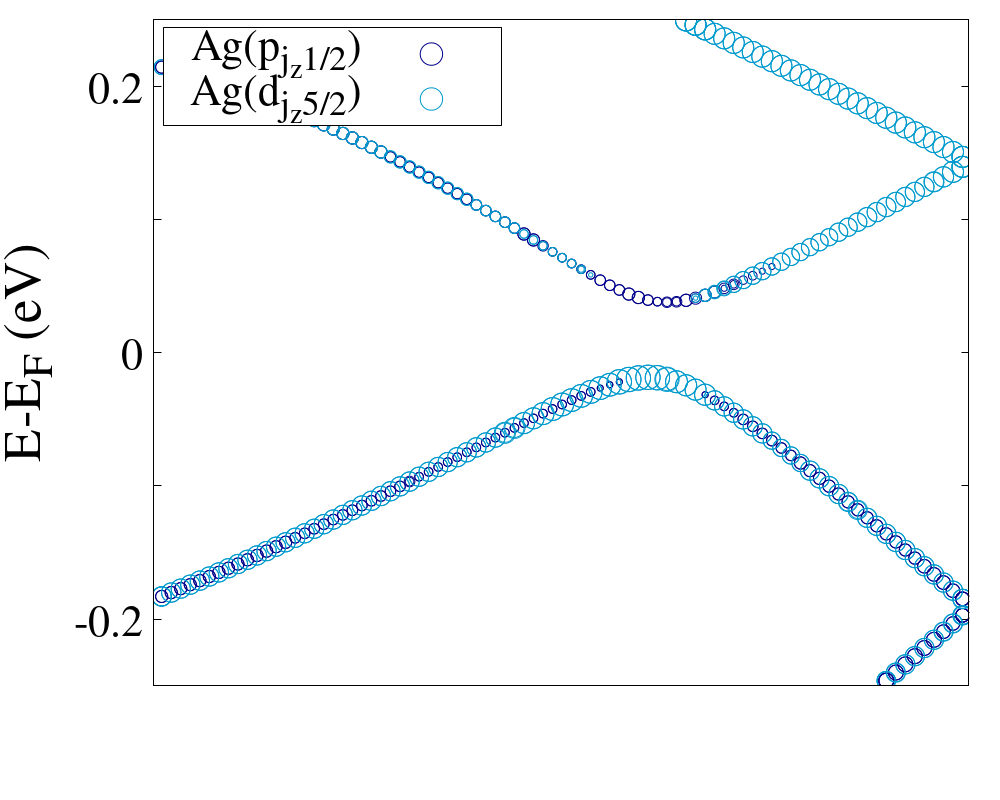}
    \caption{}
    \end{subfigure}
    \caption{One-dimensional Ag$_2$Se$_2$ (Ag: silver; Se: green) after reconstruction: (a) electronic band structure and PDOS. Large contributions at E$_F$ come from Se-\emph{p$_x$} and Se-\emph{p$_z$}, although an extra Se-\emph{p$_y$} term appears that was not present, or marginally so, in the undistorted unstable case (more information in SI) (b) Stable phonon dispersions along the Brillouin zone. (c) Band structures with and without SOC, where a gap opening is visible, and (d) highlight of the band inversion with SOC ($\pm j_z$ are identical).}
    \label{ag2se2}
\end{figure}

Ag$_2$Se$_2$ (shown in Fig. \ref{ag2se2}) is also interesting and unusual for two reasons. First, it is the only material in the study that exhibit instabilities in the phonon spectra of the initial phase but does not open a gap in the stable double supercell, where a Dirac-cone at the Fermi surface appears, classifying it as semi-metal.
In addition, the Dirac cone splits after the inclusion of SOC, opening a small gap, (Fig. \ref{ag2se2} (c) and (d)).  Band inversion occurs between Ag-(\emph{pj$_{1/2}$j$_{z\pm1/2}$}) and Ag-(\emph{dj$_{5/2}$j$_{z\pm5/2}$}), suggesting the system as a possible one-dimensional topological insulator \citep{jin20201d,liu2022ta}. 
To date, materials where the formation zero-energy states at the border of a nanowire, as for example Majorana edge states, occurs intrinsically are very rare; in this regard, Ag$_2$Se$_2$ as a one-dimensional Van-der-Waals exfoliable wire could be promising to extend the family of pure 1D topological materials.

\begin{figure}[h!]
    \centering
    \begin{subfigure}{1\textwidth}
    \centering
    \includegraphics[scale=0.20]{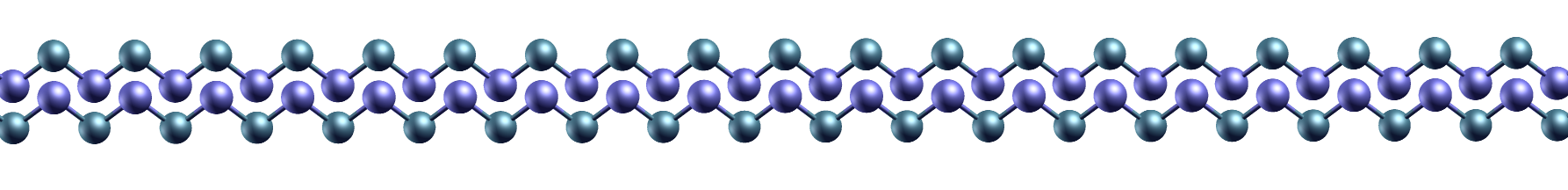}
    \end{subfigure}
    \vspace{0.3cm}
    \begin{subfigure}{0.45\textwidth}
    \centering
    \includegraphics[scale=0.21]{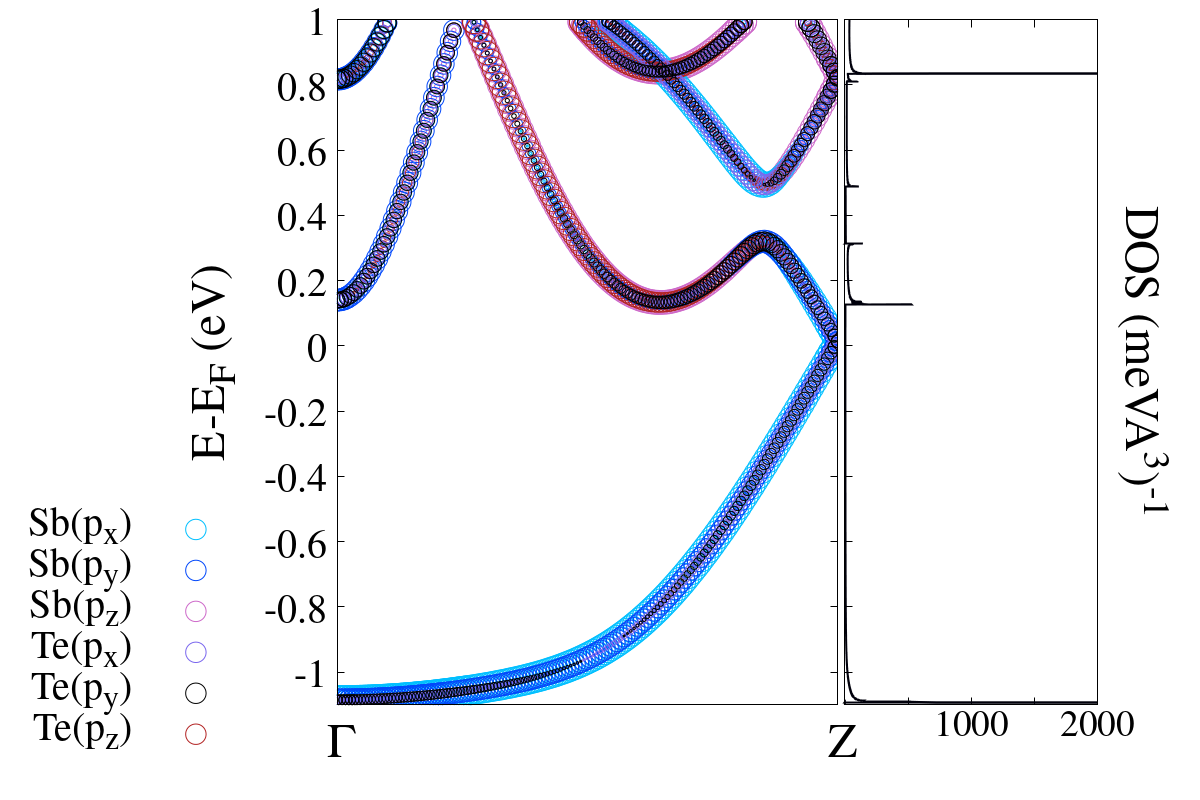}
    \caption{}
    \end{subfigure} \hspace{1cm}
    \begin{subfigure}{0.45\textwidth}
    \includegraphics[scale=0.20]{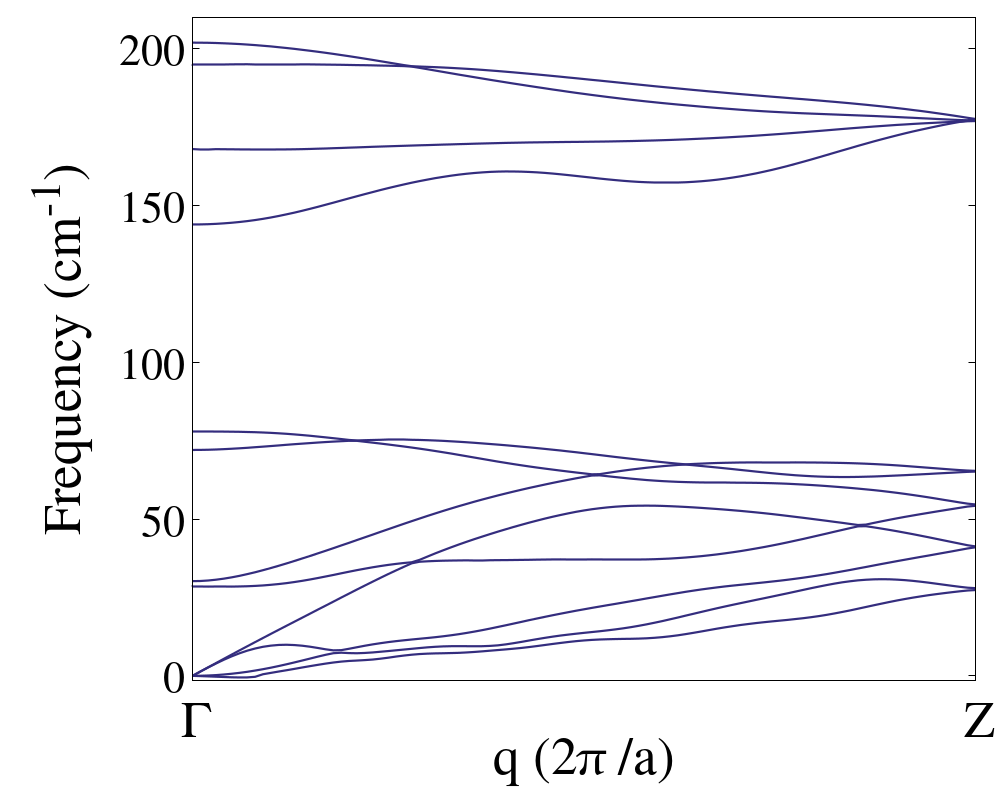}
    \caption{}
    \end{subfigure}
    \caption{One-dimensional Sb$_2$Te$_2$ (Te: green, Sb: purple): (a) electronic band structure with projected density of states; the Dirac cone is visible at the edge \emph{\textbf{Z}}. Sb- and Te- \emph{p$_x$}, \emph{p$_y$} orbitals appear as the most prominent around the Fermi energy. (b) Stable phonon dispersions along the Brillouin zone.}
    \label{sb2te2}
\end{figure}

Sb$_2$Te$_2$ (shown in Fig. \ref{sb2te2}) is another novel material that emerges from the search and that is stable without reconstruction.
As can be seen from the electronic bands, this material, as Ag$_2$Se$_2$, presents a Dirac-cone on the edge of the BZ, with the Dirac point exactly at E$_F$. The Dirac cone is moreover robust against spin-orbit coupling (SOC) (see Supporting Information for comparison of the band structures with and without SOC), which rules out the system from being a possible one-dimensional topological insulator. As a zero-gap semimetal, nonetheless, Sb$_2$Te$_2$ is an interesting candidate to study excitonic insulators\citep{Varsano2017,varsano2020monolayer}, the exotic state of matter predicted fifty years ago\citep{jerome1967excitonic} but remaining elusive to this day.
We estimate the Fermi velocity around the Dirac point to be $v_F=\frac{1}{\hbar}\frac{\partial E}{\partial k}$ $\sim$ 5.0 x $10^5$ m/s, comparable with other one-dimensional materials like SWCNTs (8.1 x $10^5$ m/s)\citep{tans1997individual} and 2D materials such as graphene (10 x $10^5$ m/s)\citep{geim2009graphene}.
As with previous cases, the projected density of states shows that around the Fermi energy there are mostly contributions from orbitals living on the plane perpendicular to the direction of the chain (both Sb- and Te- p$_x$, p$_y$).
This characteristic is shared among the four metallic/semi-metallic stable wires; as discussed in the previous section the character of the wavefunctions around the Fermi energy\citep{Derriche2022Suppression} may play a role in the stability of these wires. 

\begin{figure}[h!]
    \centering
    \begin{subfigure}{1\textwidth}
    \centering
    \includegraphics[scale=0.20]{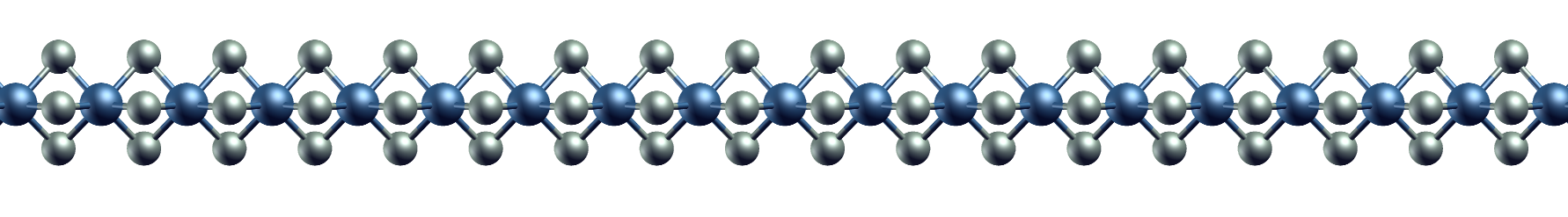}
    \end{subfigure}
    \vspace{0.3cm}
    \begin{subfigure}{0.45\textwidth}
    \centering
    \includegraphics[scale=0.21]{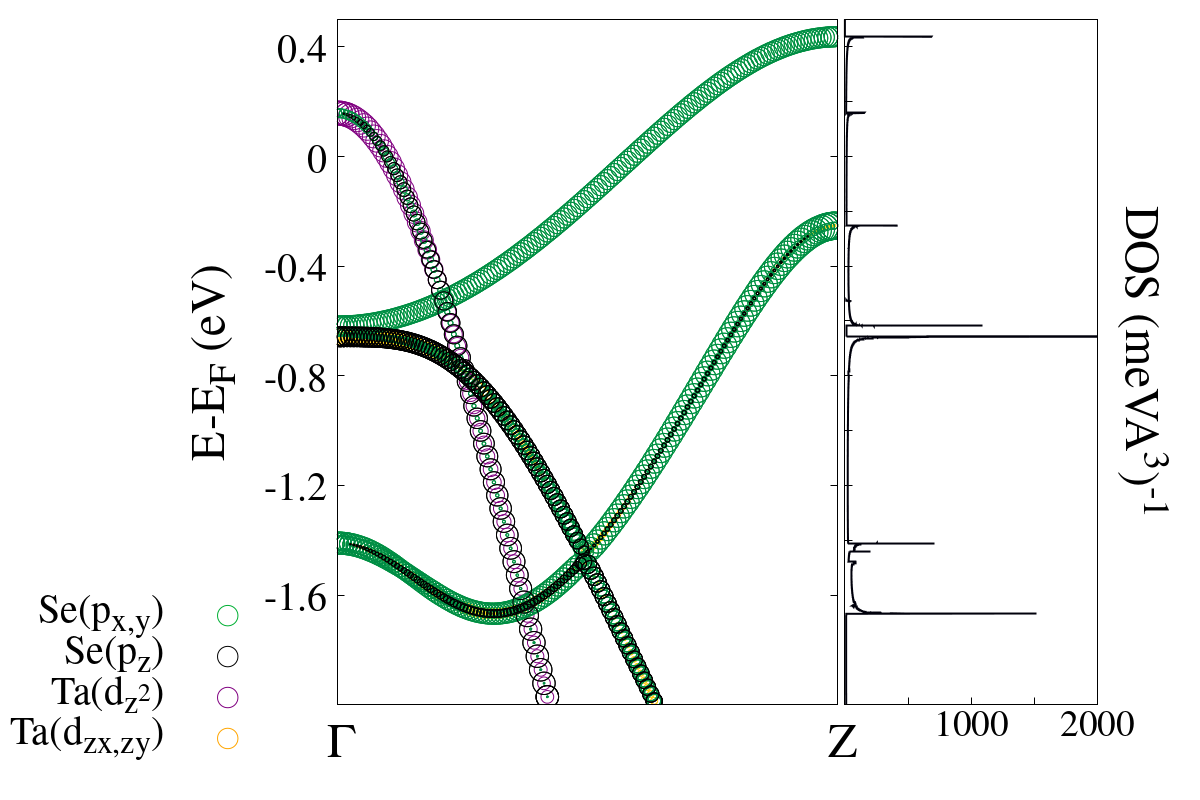}
    \caption{}
    \end{subfigure} \hspace{1cm}
    \begin{subfigure}{0.45\textwidth}
    \includegraphics[scale=0.20]{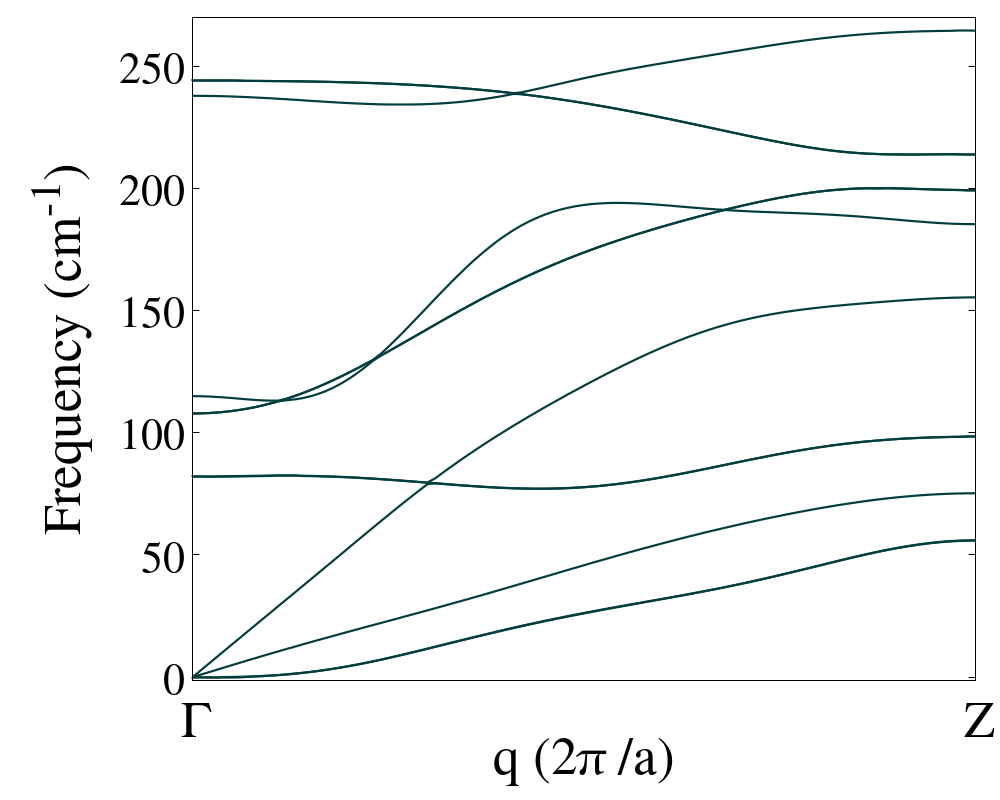}
    \caption{}
    \end{subfigure}
    \caption{One-dimensional TaSe$_3$ (Ta: blue, Se: grey): (a) electronic band structure with projected density of states. At the Fermi surface we observe contributions from Se-p$_x$, Se-p$_y$ orbitals on the plane perpendicular to the chain length, with a smaller Ta-d$_{z^2}$ contribution along the chain. (b) Stable phonon dispersions along the Brillouin zone.}
    \label{tase3}
\end{figure}

Finally, we present our results for TaSe$_3$, a well-known one-dimensional compound in the literature that we explicitly introduce in our screening as a benchmark. Since the 1970s the three-dimensional counterpart has been investigated within the family of V$^b$ transition-metal trichalcogenides (MX$_3$) \citep{yamamoto1978superconducting,nagata1989superconductivity}, mostly in the context of superconductivity. In 2016 the successful experimental exfoliation in thin nanowires \citep{stolyarov2016breakdown} proved that its wires can be remarkable good candidates for local interconnects, with a breakdown current up to 10$^8$ A/cm$^{2}$\citep{liu2017low,empante2019low}.
To the best of our knowledge, the single isolated chain has not yet been investigated. Here, we report the electronic band structure and phonon dispersions of the single TaSe$_3$ wire (Fig. \ref{tase3}). 
As mentioned, the wire is stable as is in the configuration extracted from the 3D parent (and then relaxed), Fig. \ref{tase3} (b). 
\clearpage
\section{Conclusions}
In this work, we characterise one-dimensional wires harvested from a high-throughput screening of weakly-bonded three-dimensional crystals, searching for stable metallic systems.
We discover one stable metallic atomic chain to add to the already-known TaSe$_3$, and two novel stable semi-metallic wires. 

The most promising material is CuC$_2$, representing possibly the thinnest viable metallic wire. CuC$_2$ is a straight-line chain that can be exfoliated from three distinct experimentally known 3D Van der Waals parents, NaCuC$_2$, KCuC$_2$ and RbCuC$_2$, with one the lowest exfoliation energies in the selected portfolio. CuC$_2$ exhibits the highest DOS$\big|_{E_F}$ among the metallic wires and possesses the highest Young modulus among all; the wire is bendable while preserving its metallic properties, providing also an interesting candidate for flexible electronics, ans is not attracted by O$_2$.

In addition, we identify the two stable semi-metallic Sb$_2$Te$_2$ and Ag$_2$Se$_2$ with Dirac cones at the Fermi energy. Ag$_2$Se$_2$ shows a small gap opening after the inclusion of the spin-orbit effects (SOC) and band inversion between Ag-\emph{p} and Ag\emph{d}, suggesting a potential 1D topological insulator.
Sb$_2$Te$_2$ has a robust Dirac-cone against SOC with a Fermi velocity of 5.0 x $10^5$, m/s comparable with other one-dimensional materials like SWCNTs and 2D materials as graphene.

With this study we also provide a comprehensive scenario of the different behaviour of Peierls instabilities in realistic one-dimensional materials; the resulting systems can serve as a powerful playground for fundamental research in this emergent field.

\section{Methods}\label{methods}
All the density-functional theory and density-functional perturbation theory calculations reported in this work are performed using the open-source Quantum ESPRESSO distribution\citep{giannozzi2009quantum}. For the one-dimensional wires we employ the PBE functional \citep{PBE} for the exchange-correlation term and pseudopotentials (with suggested kinetic energy cut-off) from the SSSP v1.1 PBE efficiency \citep{prandini2018precision} (for fully relativistic pseudopotentials we use the PSlibrary1.0.0.US from Ref. \citenum{dal2014pseudopotentials}; for the bulk calculations we employ the Van-der-Waals functional VdW-DF2-C09 and pseudopotentials from the SSSP v1.0 PBE efficiency).
The Brillouin zone is sampled with a \textit{k}-points distance of at least 0.09 \AA$^{-1}$ and a \emph{cold} smearing \citep{marzari1999thermal} of 0.02 Ry. For detailed density-of-states and band structures of stable metallic materials a smearing of 0.01 Ry and a 1x1x300 \textit{k}-point sampling (at least 0.006 \AA$^{-1}$) is used. The unit cells have $\sim$ 25 \r{A} of vacuum in the x-y plane transverse to the direction of the wire to remove interactions with the periodic images\citep{kozinsky2006static}.

Phonons have been computed with DFPT \citep{baroni2001phonons} after a tight relaxation, until forces on each atom were less than at least 10$^{-4}$ Ry/au. The \textbf{q}-point grid used for phonon sampling is 1x1x10. We employ the acoustic sum rule as implemented in Ref. \citenum{lin2022general} to ensure the correct long-wavelength behaviour of the acoustic phonons. 

In order to find the stable reconstructed supercells we impose a distortion from the initial positions following the lowest unstable phonon mode:
\begin{equation} \label{eta}
    \textbf{d}_{lk}=\eta \hspace{1mm}\textbf{u}_{lk}=\eta \hspace{1mm} \bm{\epsilon}_{\alpha l k} e^{i\textbf{q}\cdot \textbf{R}_{l}};
\end{equation}
where $\eta$ is an overall amplitude and $\bm{\epsilon}$ is the polarization vector describing the movements on the $\alpha$-direction (x,y,z) for the \emph{k}-th atom in the \emph{l}-th cell of the supercell. The displacement pattern is enforced on a supercell according to the corresponding phonon wave vector \textbf{q}.

The Young modulus is calculated as quadratic coefficient following the definition:
\begin{equation}\label{Y}
           Y=  \frac {\sigma (\varepsilon)} {\varepsilon}
        =\frac{F/A}{\Delta z/ z_0} = \frac{1}{V_0}\frac{\partial^2 E}{\partial \varepsilon ^2};
    \end{equation}
where $\sigma$ is the tensile stress in response of a strain applied along the $\hat z$ direction. The strain $\varepsilon$ is an adimensional parameter defined as $\Delta z/z_0$, where $z_0$ is the length of the chain. $V_0$ is the quantum volume defined in Ref. \citenum{cococcioni2005electronic} computed with the Quantum ENVIRON package\citep{andreussi2012revised} as a linear coefficient of the PV energy term, (with different values of environment pressure 1, 0.1, 0.01, 0.001,
0.0001 GPa). To do this, we insert the system in a vacuum environment with the static permittivity set to $\varepsilon$=1.0000001. In order to have a comparison between our method and the results in literature, we compute the Young modulus of a carbon chain in the polyyne structure (the insulating carbon chain with bond alternation). The Young modulus obtained is $Y_{polyyne}$=605.6 GPa, to compare with 760 GPa for Ref. \citenum{zhang2011one}.
Finally we can obtain the wire radii as $\sqrt{V_0/\pi z}$. All the density-of-states are normalized by V$_0$.

To inspect possible magnetic ground state in the stable metallic and semi-metallic materials we relax the system giving a starting magnetization in different magnetic configurations, with spins aligned parallel (in-line) and transversal (out-of-line) to the direction of the chains, and for both the ferromagnetic and antiferromagnetic cases, using collinear calculations (npin=2). More details are reported in the Supporting Information.
In particular we check carefully the CuC$_2$ wire, performing also noncollinear calculation to account for possible spin rotations. Recent theoretical work describes the wire as half-metal\citep{min2022half} with a ferromagnetic ground-state.
Based on our findings, this magnetism is present only when performing calculations with a sparse sampling of the Brillouin zone, specifically 1x1x5\citep{min2022half}, that disappears with more careful calculations (here, 1x1x60). 

For the calculation of strained and bent CuC$_2$, we apply first a strain of \emph{n}\% with respect to the original wire length z$_0$ within a 8-cell supercell, and then we modulate the y-coordinates of the system by a sinusoidal wave:
\begin{equation}
    y' = \eta \hspace{1mm} \sin(z'\frac{2\pi}{z_0})
\end{equation}
where we set $\eta$ to 1~\AA. We then relax the system without applying any constrains on the atomic positions.
To investigate the oxidation of CuC$_2$ we perform relaxations of the system within a 9-atom cell of the linear chain in the presence of one O$_2$ molecule, starting from different random configurations and distance from the chain ranging between 1.8~\AA~and 4~\AA. We include in the calculation the  rVV10\citep{sabatini2013nonlocal} Van-der-Waals correction and we provide a starting magnetization to both Cu and O atoms.
For the hybrid functional calculations we use HSE06\citep{heyd2003hybrid} with a q-mesh for the Fock operator 1x1x40 (and, in PWscf, exxdiv\textunderscore treatment=vcut\_ws and ecutvcut=3 to treat the Coulomb-potential divergencies at small q vectors) and norm-conserving pseudopotentials from pseudoDojo\citep{van2018pseudodojo}.

All the relevant input and output files are available on the Materials Cloud\citep{talirz2020materials,cignarella2024cloud}.

\begin{acknowledgement}
This research was supported by the NCCR MARVEL, a National Centre of Competence in Research, funded by the Swiss National Science Foundation (grant number 205602).
We thank S. Grillo and L. Bastonero for fruitful and useful discussions. The authors acknowledge the Swiss National Supercomputing Centre (CSCS) for simulation time under project ID mr33.

\noindent\textit{Conflict of Interest}: The authors declare no competing financial interest.

\noindent\textbf{Supporting information available} Exfoliation energies and quantum radii. Phonon dispersions before and after the relaxation/reconstruction process and, for the stable structures, band structures, density-of-states and energy-versus-strain computed for the elastic properties. Details on the magnetic calculations for the stable metals/semi-metals. Band structures of CuC$_2$ under linear strains. Calculation of CuC$_2$ oxidation reactivity. Comparison of CuC$_2$ band structures using the hybrid functional HSE06 and semi-local PBE. Comparison of Sb$_2$Te$_2$ and TaSe$_3$ band structures with and without SOC.
\end{acknowledgement}
\bibliography{library}

\providecommand{\latin}[1]{#1}
\makeatletter
\providecommand{\doi}
  {\begingroup\let\do\@makeother\dospecials
  \catcode`\{=1 \catcode`\}=2 \doi@aux}
\providecommand{\doi@aux}[1]{\endgroup\texttt{#1}}
\makeatother
\providecommand*\mcitethebibliography{\thebibliography}
\csname @ifundefined\endcsname{endmcitethebibliography}  {\let\endmcitethebibliography\endthebibliography}{}
\begin{mcitethebibliography}{124}
\providecommand*\natexlab[1]{#1}
\providecommand*\mciteSetBstSublistMode[1]{}
\providecommand*\mciteSetBstMaxWidthForm[2]{}
\providecommand*\mciteBstWouldAddEndPuncttrue
  {\def\EndOfBibitem{\unskip.}}
\providecommand*\mciteBstWouldAddEndPunctfalse
  {\let\EndOfBibitem\relax}
\providecommand*\mciteSetBstMidEndSepPunct[3]{}
\providecommand*\mciteSetBstSublistLabelBeginEnd[3]{}
\providecommand*\EndOfBibitem{}
\mciteSetBstSublistMode{f}
\mciteSetBstMaxWidthForm{subitem}{(\alph{mcitesubitemcount})}
\mciteSetBstSublistLabelBeginEnd
  {\mcitemaxwidthsubitemform\space}
  {\relax}
  {\relax}

\bibitem[Iijima(1991)]{iijima1991helical}
Iijima,~S. Helical Microtubules of Graphitic Carbon. \emph{Nature} \textbf{1991}, \emph{354}, 56--58\relax
\mciteBstWouldAddEndPuncttrue
\mciteSetBstMidEndSepPunct{\mcitedefaultmidpunct}
{\mcitedefaultendpunct}{\mcitedefaultseppunct}\relax
\EndOfBibitem
\bibitem[Iijima and Ichihashi(1993)Iijima, and Ichihashi]{iijima1993single}
Iijima,~S.; Ichihashi,~T. Single-Shell Carbon Nanotubes of 1-nm Diameter. \emph{Nature} \textbf{1993}, \emph{363}, 603--605\relax
\mciteBstWouldAddEndPuncttrue
\mciteSetBstMidEndSepPunct{\mcitedefaultmidpunct}
{\mcitedefaultendpunct}{\mcitedefaultseppunct}\relax
\EndOfBibitem
\bibitem[Ajayan and Ebbesen(1997)Ajayan, and Ebbesen]{ajayan1997nanometre}
Ajayan,~P.; Ebbesen,~T. Nanometre-Size Tubes of Carbon. \emph{Rep. Prog. Phys.} \textbf{1997}, \emph{60}, 1025\relax
\mciteBstWouldAddEndPuncttrue
\mciteSetBstMidEndSepPunct{\mcitedefaultmidpunct}
{\mcitedefaultendpunct}{\mcitedefaultseppunct}\relax
\EndOfBibitem
\bibitem[Tans \latin{et~al.}(1997)Tans, Devoret, Dai, Thess, Smalley, Geerligs, and Dekker]{tans1997individual}
Tans,~S.~J.; Devoret,~M.~H.; Dai,~H.; Thess,~A.; Smalley,~R.~E.; Geerligs,~L.; Dekker,~C. Individual Single-Wall Carbon Nanotubes as Quantum Wires. \emph{Nature} \textbf{1997}, \emph{386}, 474--477\relax
\mciteBstWouldAddEndPuncttrue
\mciteSetBstMidEndSepPunct{\mcitedefaultmidpunct}
{\mcitedefaultendpunct}{\mcitedefaultseppunct}\relax
\EndOfBibitem
\bibitem[Kim \latin{et~al.}(2001)Kim, Shi, Majumdar, and McEuen]{kim2001thermal}
Kim,~P.; Shi,~L.; Majumdar,~A.; McEuen,~P.~L. Thermal Transport Measurements of Individual Multiwalled Nanotubes. \emph{Phys. Rev. Lett.} \textbf{2001}, \emph{87}, 215502\relax
\mciteBstWouldAddEndPuncttrue
\mciteSetBstMidEndSepPunct{\mcitedefaultmidpunct}
{\mcitedefaultendpunct}{\mcitedefaultseppunct}\relax
\EndOfBibitem
\bibitem[Geremew \latin{et~al.}(2018)Geremew, Bloodgood, Aytan, Woo, Corber, Liu, Bozhilov, Salguero, Rumyantsev, Rao, and Balandin]{geremew2018current}
Geremew,~A.; Bloodgood,~M.~A.; Aytan,~E.; Woo,~B. W.~K.; Corber,~S.~R.; Liu,~G.; Bozhilov,~K.; Salguero,~T.~T.; Rumyantsev,~S.; Rao,~M.~P.; Balandin,~A.~A. Current Carrying Capacity of Quasi-1D ZrTe3 Van der Waals Nanoribbons. \emph{IEEE Electron Device Lett.} \textbf{2018}, \emph{39}, 735--738\relax
\mciteBstWouldAddEndPuncttrue
\mciteSetBstMidEndSepPunct{\mcitedefaultmidpunct}
{\mcitedefaultendpunct}{\mcitedefaultseppunct}\relax
\EndOfBibitem
\bibitem[Stolyarov \latin{et~al.}(2016)Stolyarov, Liu, Bloodgood, Aytan, Jiang, Samnakay, Salguero, Nika, Rumyantsev, Shur, and et~al.]{stolyarov2016breakdown}
Stolyarov,~M.~A.; Liu,~G.; Bloodgood,~M.~A.; Aytan,~E.; Jiang,~C.; Samnakay,~R.; Salguero,~T.~T.; Nika,~D.~L.; Rumyantsev,~S.~L.; Shur,~M.~S.; et~al. Breakdown current density in h-BN-capped quasi-1D TaSe3 metallic nanowires: prospects of interconnect applications. \emph{Nanoscale} \textbf{2016}, \emph{8}, 15774--15782\relax
\mciteBstWouldAddEndPuncttrue
\mciteSetBstMidEndSepPunct{\mcitedefaultmidpunct}
{\mcitedefaultendpunct}{\mcitedefaultseppunct}\relax
\EndOfBibitem
\bibitem[Park \latin{et~al.}(2020)Park, Hwang, Kim, and Park]{Park2020}
Park,~J.; Hwang,~J.~C.; Kim,~G.~G.; Park,~J.-U. Flexible Electronics Based on One-Dimensional and Two-Dimensional Hybrid Nanomaterials. \emph{InfoMat} \textbf{2020}, \emph{2}, 33--56\relax
\mciteBstWouldAddEndPuncttrue
\mciteSetBstMidEndSepPunct{\mcitedefaultmidpunct}
{\mcitedefaultendpunct}{\mcitedefaultseppunct}\relax
\EndOfBibitem
\bibitem[Gruner(2018)]{gruner2018density}
Gruner,~G. \emph{Density Waves in Solids}; CRC press, 2018\relax
\mciteBstWouldAddEndPuncttrue
\mciteSetBstMidEndSepPunct{\mcitedefaultmidpunct}
{\mcitedefaultendpunct}{\mcitedefaultseppunct}\relax
\EndOfBibitem
\bibitem[Peierls and Peierls(1955)Peierls, and Peierls]{peierls1955quantum}
Peierls,~R.; Peierls,~R.~E. \emph{Quantum Theory of Solids}; Oxford University Press, 1955\relax
\mciteBstWouldAddEndPuncttrue
\mciteSetBstMidEndSepPunct{\mcitedefaultmidpunct}
{\mcitedefaultendpunct}{\mcitedefaultseppunct}\relax
\EndOfBibitem
\bibitem[Luttinger(1963)]{luttinger1963exactly}
Luttinger,~J. An Exactly Soluble Model of a Many-Fermion System. \emph{J. Math. Phys.} \textbf{1963}, \emph{4}, 1154--1162\relax
\mciteBstWouldAddEndPuncttrue
\mciteSetBstMidEndSepPunct{\mcitedefaultmidpunct}
{\mcitedefaultendpunct}{\mcitedefaultseppunct}\relax
\EndOfBibitem
\bibitem[Haldane(1981)]{haldane1981luttinger}
Haldane,~F. 'Luttinger liquid theory' of One-Dimensional Quantum Fluids. I. Properties of the Luttinger Model and Their Extension to the General 1D Interacting Spinless Fermi Gas. \emph{J. Phys. C: Solid State Phys.} \textbf{1981}, \emph{14}, 2585\relax
\mciteBstWouldAddEndPuncttrue
\mciteSetBstMidEndSepPunct{\mcitedefaultmidpunct}
{\mcitedefaultendpunct}{\mcitedefaultseppunct}\relax
\EndOfBibitem
\bibitem[Voit(1995)]{voit1995one}
Voit,~J. One-Dimensional Fermi Liquids. \emph{Rep. Prog. Phys.} \textbf{1995}, \emph{58}, 977\relax
\mciteBstWouldAddEndPuncttrue
\mciteSetBstMidEndSepPunct{\mcitedefaultmidpunct}
{\mcitedefaultendpunct}{\mcitedefaultseppunct}\relax
\EndOfBibitem
\bibitem[Varsano \latin{et~al.}(2017)Varsano, Sorella, Sangalli, Barborini, Corni, Molinari, and Rontani]{Varsano2017}
Varsano,~D.; Sorella,~S.; Sangalli,~D.; Barborini,~M.; Corni,~S.; Molinari,~E.; Rontani,~M. Carbon Nanotubes as Excitonic Insulators. \emph{Nat. Commun.} \textbf{2017}, \emph{8}, 1461\relax
\mciteBstWouldAddEndPuncttrue
\mciteSetBstMidEndSepPunct{\mcitedefaultmidpunct}
{\mcitedefaultendpunct}{\mcitedefaultseppunct}\relax
\EndOfBibitem
\bibitem[Gambardella \latin{et~al.}(2002)Gambardella, Dallmeyer, Maiti, Malagoli, Eberhardt, Kern, and Carbone]{gambardella2002ferromagnetism}
Gambardella,~P.; Dallmeyer,~A.; Maiti,~K.; Malagoli,~M.; Eberhardt,~W.; Kern,~K.; Carbone,~C. Ferromagnetism in One-Dimensional Monatomic Metal Chains. \emph{Nature} \textbf{2002}, \emph{416}, 301--304\relax
\mciteBstWouldAddEndPuncttrue
\mciteSetBstMidEndSepPunct{\mcitedefaultmidpunct}
{\mcitedefaultendpunct}{\mcitedefaultseppunct}\relax
\EndOfBibitem
\bibitem[Crain \latin{et~al.}(2004)Crain, McChesney, Zheng, Gallagher, Snijders, Bissen, Gundelach, Erwin, and Himpsel]{crain2004chains}
Crain,~J.; McChesney,~J.; Zheng,~F.; Gallagher,~M.; Snijders,~P.; Bissen,~M.; Gundelach,~C.; Erwin,~S.~C.; Himpsel,~F. Chains of Gold Atoms with Tailored Electronic States. \emph{Phys. Rev. B} \textbf{2004}, \emph{69}, 125401\relax
\mciteBstWouldAddEndPuncttrue
\mciteSetBstMidEndSepPunct{\mcitedefaultmidpunct}
{\mcitedefaultendpunct}{\mcitedefaultseppunct}\relax
\EndOfBibitem
\bibitem[Zeng \latin{et~al.}(2008)Zeng, Kent, Kim, Li, and Weitering]{zeng2008charge}
Zeng,~C.; Kent,~P.; Kim,~T.-H.; Li,~A.-P.; Weitering,~H.~H. Charge-Order Fluctuations in One-Dimensional Silicides. \emph{Nat. Mater.} \textbf{2008}, \emph{7}, 539--542\relax
\mciteBstWouldAddEndPuncttrue
\mciteSetBstMidEndSepPunct{\mcitedefaultmidpunct}
{\mcitedefaultendpunct}{\mcitedefaultseppunct}\relax
\EndOfBibitem
\bibitem[Ferstl \latin{et~al.}(2016)Ferstl, Hammer, Sobel, Gubo, Heinz, Schneider, Mittendorfer, and Redinger]{ferstl2016self}
Ferstl,~P.; Hammer,~L.; Sobel,~C.; Gubo,~M.; Heinz,~K.; Schneider,~M.~A.; Mittendorfer,~F.; Redinger,~J. Self-Organized Growth, Structure, and Magnetism of Monatomic Transition-Metal Oxide Chains. \emph{Phys. Rev. Lett.} \textbf{2016}, \emph{117}, 046101\relax
\mciteBstWouldAddEndPuncttrue
\mciteSetBstMidEndSepPunct{\mcitedefaultmidpunct}
{\mcitedefaultendpunct}{\mcitedefaultseppunct}\relax
\EndOfBibitem
\bibitem[Qin \latin{et~al.}(2018)Qin, Qiu, He, Jian, Si, Duan, Charnas, Zemlyanov, Wang, Shao, and et~al.]{qin2018epitaxial}
Qin,~J.-K.; Qiu,~G.; He,~W.; Jian,~J.; Si,~M.-W.; Duan,~Y.-Q.; Charnas,~A.; Zemlyanov,~D.~Y.; Wang,~H.-Y.; Shao,~W.-Z.; et~al. Epitaxial Growth of 1D Atomic Chain Based Se Nanoplates on Monolayer ReS2 for High-Performance Photodetectors. \emph{Adv. Funct. Mater.} \textbf{2018}, \emph{28}, 1806254\relax
\mciteBstWouldAddEndPuncttrue
\mciteSetBstMidEndSepPunct{\mcitedefaultmidpunct}
{\mcitedefaultendpunct}{\mcitedefaultseppunct}\relax
\EndOfBibitem
\bibitem[Guo \latin{et~al.}(2022)Guo, Fu, Zhang, Zhu, Yao, Xu, An, Wang, Tang, Deng, and et~al.]{guo2022direct}
Guo,~S.; Fu,~J.; Zhang,~P.; Zhu,~C.; Yao,~H.; Xu,~M.; An,~B.; Wang,~X.; Tang,~B.; Deng,~Y.; et~al. Direct Growth of Single-Metal-Atom Chains. \emph{Nat. Synth.} \textbf{2022}, \emph{1}, 245--253\relax
\mciteBstWouldAddEndPuncttrue
\mciteSetBstMidEndSepPunct{\mcitedefaultmidpunct}
{\mcitedefaultendpunct}{\mcitedefaultseppunct}\relax
\EndOfBibitem
\bibitem[Sanna \latin{et~al.}(2018)Sanna, Lichtenstein, Mamiyev, Tegenkamp, and Pfnür]{sanna2018one}
Sanna,~S.; Lichtenstein,~T.; Mamiyev,~Z.; Tegenkamp,~C.; Pfnür,~H. How One-Dimensional Are Atomic Gold Chains on a Substrate? \emph{J. Phys. Chem. C} \textbf{2018}, \emph{122}, 25580--25588\relax
\mciteBstWouldAddEndPuncttrue
\mciteSetBstMidEndSepPunct{\mcitedefaultmidpunct}
{\mcitedefaultendpunct}{\mcitedefaultseppunct}\relax
\EndOfBibitem
\bibitem[Yogi \latin{et~al.}(2022)Yogi, Koch, Sanna, and Pfn{\"u}r]{yogi2022electronic}
Yogi,~P.; Koch,~J.; Sanna,~S.; Pfn{\"u}r,~H. Electronic Phase Transitions in Quasi-One-Dimensional Atomic Chains: Au Wires on Si (553). \emph{Phys. Rev. B} \textbf{2022}, \emph{105}, 235407\relax
\mciteBstWouldAddEndPuncttrue
\mciteSetBstMidEndSepPunct{\mcitedefaultmidpunct}
{\mcitedefaultendpunct}{\mcitedefaultseppunct}\relax
\EndOfBibitem
\bibitem[Stonemeyer \latin{et~al.}(2020)Stonemeyer, Cain, Oh, Azizi, Elasha, Thiel, Song, Ercius, Cohen, and Zettl]{stonemeyer2020stabilization}
Stonemeyer,~S.; Cain,~J.~D.; Oh,~S.; Azizi,~A.; Elasha,~M.; Thiel,~M.; Song,~C.; Ercius,~P.; Cohen,~M.~L.; Zettl,~A. Stabilization of NbTe3, VTe3, and TiTe3 via Nanotube Encapsulation. \emph{J. Am. Chem. Soc.} \textbf{2020}, \emph{143}, 4563--4568\relax
\mciteBstWouldAddEndPuncttrue
\mciteSetBstMidEndSepPunct{\mcitedefaultmidpunct}
{\mcitedefaultendpunct}{\mcitedefaultseppunct}\relax
\EndOfBibitem
\bibitem[Kashtiban \latin{et~al.}(2021)Kashtiban, Burdanova, Vasylenko, Wynn, Medeiros, Ramasse, Morris, Quigley, Lloyd-Hughes, and Sloan]{kashtiban2021linear}
Kashtiban,~R.~J.; Burdanova,~M.~G.; Vasylenko,~A.; Wynn,~J.; Medeiros,~P.~V.; Ramasse,~Q.; Morris,~A.~J.; Quigley,~D.; Lloyd-Hughes,~J.; Sloan,~J. Linear and Helical Cesium Iodide Atomic Chains in Ultranarrow Single-Walled Carbon Nanotubes: Impact on Optical Properties. \emph{ACS Nano} \textbf{2021}, \emph{15}, 13389--13398\relax
\mciteBstWouldAddEndPuncttrue
\mciteSetBstMidEndSepPunct{\mcitedefaultmidpunct}
{\mcitedefaultendpunct}{\mcitedefaultseppunct}\relax
\EndOfBibitem
\bibitem[Yan \latin{et~al.}(2017)Yan, Hohman, Li, Jia, Solis-Ibarra, Wu, Dahl, Carlson, Tkachenko, Fokin, and et~al.]{yan2017hybrid}
Yan,~H.; Hohman,~J.~N.; Li,~F.~H.; Jia,~C.; Solis-Ibarra,~D.; Wu,~B.; Dahl,~J.~E.; Carlson,~R.~M.; Tkachenko,~B.~A.; Fokin,~A.~A.; et~al. Hybrid Metal--Organic Chalcogenide Nanowires with Electrically Conductive Inorganic Core Through Diamondoid-Directed Assembly. \emph{Nat. Mater.} \textbf{2017}, \emph{16}, 349--355\relax
\mciteBstWouldAddEndPuncttrue
\mciteSetBstMidEndSepPunct{\mcitedefaultmidpunct}
{\mcitedefaultendpunct}{\mcitedefaultseppunct}\relax
\EndOfBibitem
\bibitem[Xiao \latin{et~al.}(2018)Xiao, Burg, Zhou, Yan, Wang, Ding, Reed, Miller, and Dauskardt]{xiao2018electrically}
Xiao,~Q.; Burg,~J.~A.; Zhou,~Y.; Yan,~H.; Wang,~C.; Ding,~Y.; Reed,~E.; Miller,~R.~D.; Dauskardt,~R.~H. Electrically Conductive Copper Core--Shell Nanowires Through Benzenethiol-Directed Assembly. \emph{Nano Lett.} \textbf{2018}, \emph{18}, 4900--4907\relax
\mciteBstWouldAddEndPuncttrue
\mciteSetBstMidEndSepPunct{\mcitedefaultmidpunct}
{\mcitedefaultendpunct}{\mcitedefaultseppunct}\relax
\EndOfBibitem
\bibitem[Balandin \latin{et~al.}(2022)Balandin, Kargar, Salguero, and Lake]{balandin2022one}
Balandin,~A.~A.; Kargar,~F.; Salguero,~T.~T.; Lake,~R.~K. One-Dimensional Van der Waals Quantum Materials. \emph{Mater. Today} \textbf{2022}, \emph{55}, 74--91\relax
\mciteBstWouldAddEndPuncttrue
\mciteSetBstMidEndSepPunct{\mcitedefaultmidpunct}
{\mcitedefaultendpunct}{\mcitedefaultseppunct}\relax
\EndOfBibitem
\bibitem[Meng \latin{et~al.}(2022)Meng, Wang, and Ho]{meng2022one}
Meng,~Y.; Wang,~W.; Ho,~J.~C. One-Dimensional Atomic Chains for Ultimate-Scaled Electronics. \emph{ACS Nano} \textbf{2022}, \emph{16}, 13314--13322\relax
\mciteBstWouldAddEndPuncttrue
\mciteSetBstMidEndSepPunct{\mcitedefaultmidpunct}
{\mcitedefaultendpunct}{\mcitedefaultseppunct}\relax
\EndOfBibitem
\bibitem[Novoselov \latin{et~al.}(2012)Novoselov, Fal'ko, Colombo, Gellert, Schwab, and Kim]{novoselov2012roadmap}
Novoselov,~K.~S.; Fal'ko,~V.~I.; Colombo,~L.; Gellert,~P.~R.; Schwab,~M.~G.; Kim,~K. A Roadmap for Graphene. \emph{Nature} \textbf{2012}, \emph{490}, 192--200\relax
\mciteBstWouldAddEndPuncttrue
\mciteSetBstMidEndSepPunct{\mcitedefaultmidpunct}
{\mcitedefaultendpunct}{\mcitedefaultseppunct}\relax
\EndOfBibitem
\bibitem[Wang \latin{et~al.}(2012)Wang, Kalantar-Zadeh, Kis, Coleman, and Strano]{wang2012electronics}
Wang,~Q.~H.; Kalantar-Zadeh,~K.; Kis,~A.; Coleman,~J.~N.; Strano,~M.~S. Electronics and Optoelectronics of Two-dimensional Transition Metal Dichalcogenides. \emph{Nat. Nanotechnol.} \textbf{2012}, \emph{7}, 699--712\relax
\mciteBstWouldAddEndPuncttrue
\mciteSetBstMidEndSepPunct{\mcitedefaultmidpunct}
{\mcitedefaultendpunct}{\mcitedefaultseppunct}\relax
\EndOfBibitem
\bibitem[Nicolosi \latin{et~al.}(2013)Nicolosi, Chhowalla, Kanatzidis, Strano, and Coleman]{nicolosi2013liquid}
Nicolosi,~V.; Chhowalla,~M.; Kanatzidis,~M.~G.; Strano,~M.~S.; Coleman,~J.~N. Liquid Exfoliation of Layered Materials. \emph{Science} \textbf{2013}, \emph{340}, 1226419\relax
\mciteBstWouldAddEndPuncttrue
\mciteSetBstMidEndSepPunct{\mcitedefaultmidpunct}
{\mcitedefaultendpunct}{\mcitedefaultseppunct}\relax
\EndOfBibitem
\bibitem[Hanlon \latin{et~al.}(2015)Hanlon, Backes, Doherty, Cucinotta, Berner, Boland, Lee, Harvey, Lynch, Gholamvand, and et~al.]{hanlon2015liquid}
Hanlon,~D.; Backes,~C.; Doherty,~E.; Cucinotta,~C.~S.; Berner,~N.~C.; Boland,~C.; Lee,~K.; Harvey,~A.; Lynch,~P.; Gholamvand,~Z.; et~al. Liquid Exfoliation of Solvent-Stabilized Few-Layer Black Phosphorus for Applications Beyond Electronics. \emph{Nat. Commun.} \textbf{2015}, \emph{6}, 8563\relax
\mciteBstWouldAddEndPuncttrue
\mciteSetBstMidEndSepPunct{\mcitedefaultmidpunct}
{\mcitedefaultendpunct}{\mcitedefaultseppunct}\relax
\EndOfBibitem
\bibitem[Island \latin{et~al.}(2017)Island, Molina-Mendoza, Barawi, Biele, Flores, Clamagirand, Ares, S{\'a}nchez, Van Der~Zant, D’Agosta, and et~al.]{island2017electronics}
Island,~J.~O.; Molina-Mendoza,~A.~J.; Barawi,~M.; Biele,~R.; Flores,~E.; Clamagirand,~J.~M.; Ares,~J.~R.; S{\'a}nchez,~C.; Van Der~Zant,~H.~S.; D’Agosta,~R.; et~al. Electronics and Optoelectronics of Quasi-1D Layered Transition Metal Trichalcogenides. \emph{2D Mater.} \textbf{2017}, \emph{4}, 022003\relax
\mciteBstWouldAddEndPuncttrue
\mciteSetBstMidEndSepPunct{\mcitedefaultmidpunct}
{\mcitedefaultendpunct}{\mcitedefaultseppunct}\relax
\EndOfBibitem
\bibitem[Lipatov \latin{et~al.}(2018)Lipatov, Loes, Lu, Dai, Patoka, Vorobeva, Muratov, Ulrich, K{ä}stner, Hoehl, and et~al.]{lipatov2018quasi}
Lipatov,~A.; Loes,~M.~J.; Lu,~H.; Dai,~J.; Patoka,~P.; Vorobeva,~N.~S.; Muratov,~D.~S.; Ulrich,~G.; K{ä}stner,~B.; Hoehl,~A.; et~al. Quasi-1D TiS3 nanoribbons: mechanical exfoliation and thickness-dependent Raman spectroscopy. \emph{ACS Nano} \textbf{2018}, \emph{12}, 12713--12720\relax
\mciteBstWouldAddEndPuncttrue
\mciteSetBstMidEndSepPunct{\mcitedefaultmidpunct}
{\mcitedefaultendpunct}{\mcitedefaultseppunct}\relax
\EndOfBibitem
\bibitem[Barani \latin{et~al.}(2021)Barani, Kargar, Ghafouri, Ghosh, Godziszewski, Baraghani, Yashchyshyn, Cywi{\'n}ski, Rumyantsev, Salguero, and et~al.]{barani2021electrically}
Barani,~Z.; Kargar,~F.; Ghafouri,~Y.; Ghosh,~S.; Godziszewski,~K.; Baraghani,~S.; Yashchyshyn,~Y.; Cywi{\'n}ski,~G.; Rumyantsev,~S.; Salguero,~T.~T.; et~al. Electrically Insulating Flexible Films with Quasi-1D van der Waals Fillers as Efficient Electromagnetic Shields in the GHz and Sub-THz Frequency Bands. \emph{Adv. Mater.} \textbf{2021}, \emph{33}, 2007286\relax
\mciteBstWouldAddEndPuncttrue
\mciteSetBstMidEndSepPunct{\mcitedefaultmidpunct}
{\mcitedefaultendpunct}{\mcitedefaultseppunct}\relax
\EndOfBibitem
\bibitem[Kargar \latin{et~al.}(2022)Kargar, Barani, Sesing, Mai, Debnath, Zhang, Liu, Zhu, Ghosh, Biacchi, and et~al.]{kargar2022elemental}
Kargar,~F.; Barani,~Z.; Sesing,~N.~R.; Mai,~T.~T.; Debnath,~T.; Zhang,~H.; Liu,~Y.; Zhu,~Y.; Ghosh,~S.; Biacchi,~A.~J.; et~al. Elemental Excitations in MoI3 One-Dimensional Van der Waals Nanowires. \emph{Appl. Phys. Lett.} \textbf{2022}, \emph{121}\relax
\mciteBstWouldAddEndPuncttrue
\mciteSetBstMidEndSepPunct{\mcitedefaultmidpunct}
{\mcitedefaultendpunct}{\mcitedefaultseppunct}\relax
\EndOfBibitem
\bibitem[Leb{\`e}gue \latin{et~al.}(2013)Leb{\`e}gue, Bj{\"o}rkman, Klintenberg, Nieminen, and Eriksson]{lebegue2013two}
Leb{\`e}gue,~S.; Bj{\"o}rkman,~T.; Klintenberg,~M.; Nieminen,~R.~M.; Eriksson,~O. Two-Dimensional Materials from Data Filtering and Ab Initio Calculations. \emph{Phys. Rev. X} \textbf{2013}, \emph{3}, 031002\relax
\mciteBstWouldAddEndPuncttrue
\mciteSetBstMidEndSepPunct{\mcitedefaultmidpunct}
{\mcitedefaultendpunct}{\mcitedefaultseppunct}\relax
\EndOfBibitem
\bibitem[Ashton \latin{et~al.}(2017)Ashton, Paul, Sinnott, and Hennig]{ashton2017topology}
Ashton,~M.; Paul,~J.; Sinnott,~S.~B.; Hennig,~R.~G. Topology-Scaling Identification of Layered Solids and Stable Exfoliated 2D Materials. \emph{Phys. Rev. Lett.} \textbf{2017}, \emph{118}, 106101\relax
\mciteBstWouldAddEndPuncttrue
\mciteSetBstMidEndSepPunct{\mcitedefaultmidpunct}
{\mcitedefaultendpunct}{\mcitedefaultseppunct}\relax
\EndOfBibitem
\bibitem[Choudhary \latin{et~al.}(2017)Choudhary, Kalish, Beams, and Tavazza]{choudhary2017high}
Choudhary,~K.; Kalish,~I.; Beams,~R.; Tavazza,~F. High-Throughput Identification and Characterization of Two-Dimensional Materials using Density Functional Theory. \emph{Scientific Reports} \textbf{2017}, \emph{7}, 1--16\relax
\mciteBstWouldAddEndPuncttrue
\mciteSetBstMidEndSepPunct{\mcitedefaultmidpunct}
{\mcitedefaultendpunct}{\mcitedefaultseppunct}\relax
\EndOfBibitem
\bibitem[Cheon \latin{et~al.}(2017)Cheon, Duerloo, Sendek, Porter, Chen, and Reed]{cheon2017data}
Cheon,~G.; Duerloo,~K.-A.~N.; Sendek,~A.~D.; Porter,~C.; Chen,~Y.; Reed,~E.~J. Data Mining for New Two-and One-Dimensional Weakly Bonded Solids and Lattice-Commensurate Heterostructures. \emph{Nano Lett.} \textbf{2017}, \emph{17}, 1915--1923\relax
\mciteBstWouldAddEndPuncttrue
\mciteSetBstMidEndSepPunct{\mcitedefaultmidpunct}
{\mcitedefaultendpunct}{\mcitedefaultseppunct}\relax
\EndOfBibitem
\bibitem[Mounet \latin{et~al.}(2018)Mounet, Gibertini, Schwaller, Campi, Merkys, Marrazzo, Sohier, Castelli, Cepellotti, Pizzi, and et~al.]{mounet2018two}
Mounet,~N.; Gibertini,~M.; Schwaller,~P.; Campi,~D.; Merkys,~A.; Marrazzo,~A.; Sohier,~T.; Castelli,~I.~E.; Cepellotti,~A.; Pizzi,~G.; et~al. Two-Dimensional Materials from High--Throughput Computational Exfoliation of Experimentally Known Compounds. \emph{Nat. Nanotechnol.} \textbf{2018}, \emph{13}, 246--252\relax
\mciteBstWouldAddEndPuncttrue
\mciteSetBstMidEndSepPunct{\mcitedefaultmidpunct}
{\mcitedefaultendpunct}{\mcitedefaultseppunct}\relax
\EndOfBibitem
\bibitem[Haastrup \latin{et~al.}(2018)Haastrup, Strange, Pandey, Deilmann, Schmidt, Hinsche, Gjerding, Torelli, Larsen, Riis-Jensen, and et~al.]{haastrup2018computational}
Haastrup,~S.; Strange,~M.; Pandey,~M.; Deilmann,~T.; Schmidt,~P.~S.; Hinsche,~N.~F.; Gjerding,~M.~N.; Torelli,~D.; Larsen,~P.~M.; Riis-Jensen,~A.~C.; et~al. The Computational 2D Materials Database: High-Throughput Modeling and Discovery of Atomically Thin Crystals. \emph{2D Mater.} \textbf{2018}, \emph{5}, 042002\relax
\mciteBstWouldAddEndPuncttrue
\mciteSetBstMidEndSepPunct{\mcitedefaultmidpunct}
{\mcitedefaultendpunct}{\mcitedefaultseppunct}\relax
\EndOfBibitem
\bibitem[Larsen \latin{et~al.}(2019)Larsen, Pandey, Strange, and Jacobsen]{larsen2019definition}
Larsen,~P.~M.; Pandey,~M.; Strange,~M.; Jacobsen,~K.~W. Definition of a Scoring Parameter to Identify Low-Dimensional Materials Components. \emph{Phys. Rev. Mater.} \textbf{2019}, \emph{3}, 034003\relax
\mciteBstWouldAddEndPuncttrue
\mciteSetBstMidEndSepPunct{\mcitedefaultmidpunct}
{\mcitedefaultendpunct}{\mcitedefaultseppunct}\relax
\EndOfBibitem
\bibitem[Campi \latin{et~al.}(2023)Campi, Mounet, Gibertini, Pizzi, and Marzari]{campi2023expansion}
Campi,~D.; Mounet,~N.; Gibertini,~M.; Pizzi,~G.; Marzari,~N. Expansion of the Materials Cloud 2D Database. \emph{ACS Nano} \textbf{2023}, \emph{17}, 11268--11278\relax
\mciteBstWouldAddEndPuncttrue
\mciteSetBstMidEndSepPunct{\mcitedefaultmidpunct}
{\mcitedefaultendpunct}{\mcitedefaultseppunct}\relax
\EndOfBibitem
\bibitem[Shang \latin{et~al.}(2020)Shang, Fu, Zhou, and Zhao]{shang2020atomic}
Shang,~C.; Fu,~L.; Zhou,~S.; Zhao,~J. Atomic Wires of Transition Metal Chalcogenides: a Family of 1D Materials for Flexible Electronics and Spintronics. \emph{JACS Au} \textbf{2020}, \emph{1}, 147--155\relax
\mciteBstWouldAddEndPuncttrue
\mciteSetBstMidEndSepPunct{\mcitedefaultmidpunct}
{\mcitedefaultendpunct}{\mcitedefaultseppunct}\relax
\EndOfBibitem
\bibitem[Zhu \latin{et~al.}(2021)Zhu, Rehn, Antoniuk, Cheon, Freitas, Krishnapriyan, and Reed]{zhu2021spectrum}
Zhu,~Y.; Rehn,~D.~A.; Antoniuk,~E.~R.; Cheon,~G.; Freitas,~R.; Krishnapriyan,~A.; Reed,~E.~J. Spectrum of Exfoliable 1D Van der Waals Molecular Wires and Their Electronic Properties. \emph{ACS Nano} \textbf{2021}, \emph{15}, 9851--9859\relax
\mciteBstWouldAddEndPuncttrue
\mciteSetBstMidEndSepPunct{\mcitedefaultmidpunct}
{\mcitedefaultendpunct}{\mcitedefaultseppunct}\relax
\EndOfBibitem
\bibitem[Moustafa \latin{et~al.}(2022)Moustafa, Larsen, Gjerding, Mortensen, Thygesen, and Jacobsen]{moustafa2022computational}
Moustafa,~H.; Larsen,~P.~M.; Gjerding,~M.~N.; Mortensen,~J.~J.; Thygesen,~K.~S.; Jacobsen,~K.~W. Computational Exfoliation of Atomically Thin One-Dimensional Materials with Application to Majorana Bound States. \emph{Phys. Rev. Mater.} \textbf{2022}, \emph{6}, 064202\relax
\mciteBstWouldAddEndPuncttrue
\mciteSetBstMidEndSepPunct{\mcitedefaultmidpunct}
{\mcitedefaultendpunct}{\mcitedefaultseppunct}\relax
\EndOfBibitem
\bibitem[Moustafa \latin{et~al.}(2023)Moustafa, Lyngby, Mortensen, Thygesen, and Jacobsen]{moustafa2023hundreds}
Moustafa,~H.; Lyngby,~P.~M.; Mortensen,~J.~J.; Thygesen,~K.~S.; Jacobsen,~K.~W. Hundreds of New, Stable, One-Dimensional Materials from a Generative Machine Learning Model. \emph{Phys. Rev. Mater.} \textbf{2023}, \emph{7}, 014007\relax
\mciteBstWouldAddEndPuncttrue
\mciteSetBstMidEndSepPunct{\mcitedefaultmidpunct}
{\mcitedefaultendpunct}{\mcitedefaultseppunct}\relax
\EndOfBibitem
\bibitem[Liu \latin{et~al.}(2017)Liu, Rumyantsev, Bloodgood, Salguero, Shur, and Balandin]{liu2017low}
Liu,~G.; Rumyantsev,~S.; Bloodgood,~M.~A.; Salguero,~T.~T.; Shur,~M.; Balandin,~A.~A. Low-Frequency Electronic Noise in Auasi-1D TaSe3 Van der Waals Nanowires. \emph{Nano Lett.} \textbf{2017}, \emph{17}, 377--383\relax
\mciteBstWouldAddEndPuncttrue
\mciteSetBstMidEndSepPunct{\mcitedefaultmidpunct}
{\mcitedefaultendpunct}{\mcitedefaultseppunct}\relax
\EndOfBibitem
\bibitem[Empante \latin{et~al.}(2019)Empante, Martinez, Wurch, Zhu, Geremew, Yamaguchi, Isarraraz, Rumyantsev, Reed, Balandin, and et~al.]{empante2019low}
Empante,~T.~A.; Martinez,~A.; Wurch,~M.; Zhu,~Y.; Geremew,~A.~K.; Yamaguchi,~K.; Isarraraz,~M.; Rumyantsev,~S.; Reed,~E.~J.; Balandin,~A.~A.; et~al. Low Resistivity and High Breakdown Current Density of 10 nm Diameter Van der Waals TaSe3 Nanowires by Chemical Vapor Deposition. \emph{Nano Lett.} \textbf{2019}, \emph{19}, 4355--4361\relax
\mciteBstWouldAddEndPuncttrue
\mciteSetBstMidEndSepPunct{\mcitedefaultmidpunct}
{\mcitedefaultendpunct}{\mcitedefaultseppunct}\relax
\EndOfBibitem
\bibitem[Desai \latin{et~al.}(2016)Desai, Madhvapathy, Sachid, Llinas, Wang, Ahn, Pitner, Kim, Bokor, Hu, and et~al.]{desai2016mos2}
Desai,~S.~B.; Madhvapathy,~S.~R.; Sachid,~A.~B.; Llinas,~J.~P.; Wang,~Q.; Ahn,~G.~H.; Pitner,~G.; Kim,~M.~J.; Bokor,~J.; Hu,~C.; et~al. MoS2 Transistors with 1-Nanometer Gate Lengths. \emph{Science} \textbf{2016}, \emph{354}, 99--102\relax
\mciteBstWouldAddEndPuncttrue
\mciteSetBstMidEndSepPunct{\mcitedefaultmidpunct}
{\mcitedefaultendpunct}{\mcitedefaultseppunct}\relax
\EndOfBibitem
\bibitem[Chalifoux and Tykwinski(2010)Chalifoux, and Tykwinski]{chalifoux2010synthesis}
Chalifoux,~W.~A.; Tykwinski,~R.~R. Synthesis of Polyynes to Model the sp-Carbon Allotrope Carbyne. \emph{Nat. Chem.} \textbf{2010}, \emph{2}, 967\relax
\mciteBstWouldAddEndPuncttrue
\mciteSetBstMidEndSepPunct{\mcitedefaultmidpunct}
{\mcitedefaultendpunct}{\mcitedefaultseppunct}\relax
\EndOfBibitem
\bibitem[Shi \latin{et~al.}(2016)Shi, Rohringer, Suenaga, Niimi, Kotakoski, Meyer, Peterlik, Wanko, Cahangirov, Rubio, and et~al.]{shi2016confined}
Shi,~L.; Rohringer,~P.; Suenaga,~K.; Niimi,~Y.; Kotakoski,~J.; Meyer,~J.~C.; Peterlik,~H.; Wanko,~M.; Cahangirov,~S.; Rubio,~A.; et~al. Confined Linear Carbon Chains As a Route to Bulk Carbyne. \emph{Nat. Mater.} \textbf{2016}, \emph{15}, 634--639\relax
\mciteBstWouldAddEndPuncttrue
\mciteSetBstMidEndSepPunct{\mcitedefaultmidpunct}
{\mcitedefaultendpunct}{\mcitedefaultseppunct}\relax
\EndOfBibitem
\bibitem[Buntov \latin{et~al.}(2019)Buntov, Zatsepin, Kitayeva, and Vagapov]{buntov2019structure}
Buntov,~E.; Zatsepin,~A.; Kitayeva,~T.; Vagapov,~A. Structure and Properties of Chained Carbon: Recent Ab Initio Studies. \emph{C} \textbf{2019}, \emph{5}, 56\relax
\mciteBstWouldAddEndPuncttrue
\mciteSetBstMidEndSepPunct{\mcitedefaultmidpunct}
{\mcitedefaultendpunct}{\mcitedefaultseppunct}\relax
\EndOfBibitem
\bibitem[La~Torre \latin{et~al.}(2015)La~Torre, Botello-Mendez, Baaziz, Charlier, and Banhart]{la2015strain}
La~Torre,~A.; Botello-Mendez,~A.; Baaziz,~W.; Charlier,~J.-C.; Banhart,~F. Strain-Induced Metal--Semiconductor Transition Observed in Atomic Carbon Chains. \emph{Nat. Commun.} \textbf{2015}, \emph{6}, 6636\relax
\mciteBstWouldAddEndPuncttrue
\mciteSetBstMidEndSepPunct{\mcitedefaultmidpunct}
{\mcitedefaultendpunct}{\mcitedefaultseppunct}\relax
\EndOfBibitem
\bibitem[Artyukhov \latin{et~al.}(2014)Artyukhov, Liu, and Yakobson]{artyukhov2014mechanically}
Artyukhov,~V.~I.; Liu,~M.; Yakobson,~B.~I. Mechanically Induced Metal--Insulator Transition in Carbyne. \emph{Nano Lett.} \textbf{2014}, \emph{14}, 4224--4229\relax
\mciteBstWouldAddEndPuncttrue
\mciteSetBstMidEndSepPunct{\mcitedefaultmidpunct}
{\mcitedefaultendpunct}{\mcitedefaultseppunct}\relax
\EndOfBibitem
\bibitem[Romanin \latin{et~al.}(2021)Romanin, Monacelli, Bianco, Errea, Mauri, and Calandra]{romanin2021dominant}
Romanin,~D.; Monacelli,~L.; Bianco,~R.; Errea,~I.; Mauri,~F.; Calandra,~M. Dominant Role of Quantum Anharmonicity in the Stability and Optical Properties of Infinite Linear Acetylenic Carbon Chains. \emph{J. Phys. Chem. Lett.} \textbf{2021}, \emph{12}, 10339--10345\relax
\mciteBstWouldAddEndPuncttrue
\mciteSetBstMidEndSepPunct{\mcitedefaultmidpunct}
{\mcitedefaultendpunct}{\mcitedefaultseppunct}\relax
\EndOfBibitem
\bibitem[Gra{\v{z}}ulis \latin{et~al.}(2012)Gra{\v{z}}ulis, Da{\v{s}}kevi{\v{c}}, Merkys, Chateigner, Lutterotti, Quiros, Serebryanaya, Moeck, Downs, and Le~Bail]{COD}
Gra{\v{z}}ulis,~S.; Da{\v{s}}kevi{\v{c}},~A.; Merkys,~A.; Chateigner,~D.; Lutterotti,~L.; Quiros,~M.; Serebryanaya,~N.~R.; Moeck,~P.; Downs,~R.~T.; Le~Bail,~A. Crystallography Open Database (COD): an open-access collection of crystal structures and platform for world-wide collaboration. \emph{Nucleic acids research} \textbf{2012}, \emph{40}, D420--D427, database version: 211196, access date: 02-01-2019\relax
\mciteBstWouldAddEndPuncttrue
\mciteSetBstMidEndSepPunct{\mcitedefaultmidpunct}
{\mcitedefaultendpunct}{\mcitedefaultseppunct}\relax
\EndOfBibitem
\bibitem[FIZ-Karlsruhe()]{ICSD}
FIZ-Karlsruhe {Inorganic Crystal Structure Database (ICSD)}. \url{http://www.fiz-karlsruhe.com/icsd.html}, database version: 2017.2, access date: 02-01-2019\relax
\mciteBstWouldAddEndPuncttrue
\mciteSetBstMidEndSepPunct{\mcitedefaultmidpunct}
{\mcitedefaultendpunct}{\mcitedefaultseppunct}\relax
\EndOfBibitem
\bibitem[Bergerhoff \latin{et~al.}(1983)Bergerhoff, Hundt, Sievers, and Brown]{bergerhoff1983inorganic}
Bergerhoff,~G.; Hundt,~R.; Sievers,~R.; Brown,~I. The Inorganic Crystal Structure Data Base. \emph{J. Chem. Inf. Comput. Sci.} \textbf{1983}, \emph{23}, 66--69\relax
\mciteBstWouldAddEndPuncttrue
\mciteSetBstMidEndSepPunct{\mcitedefaultmidpunct}
{\mcitedefaultendpunct}{\mcitedefaultseppunct}\relax
\EndOfBibitem
\bibitem[mpd()]{mpds}
The Pauling File exposed through the Materials Platform for Data Science. \url{https://mpds.io/}, database version: 1.0.0, access date: 02-01-2019\relax
\mciteBstWouldAddEndPuncttrue
\mciteSetBstMidEndSepPunct{\mcitedefaultmidpunct}
{\mcitedefaultendpunct}{\mcitedefaultseppunct}\relax
\EndOfBibitem
\bibitem[Villars \latin{et~al.}(1998)Villars, Onodera, and Iwata]{villars1998linus}
Villars,~P.; Onodera,~N.; Iwata,~S. The Linus Pauling file (LPF) and its Application to Materials Design. \emph{J. Alloys Compd.} \textbf{1998}, \emph{279}, 1--7\relax
\mciteBstWouldAddEndPuncttrue
\mciteSetBstMidEndSepPunct{\mcitedefaultmidpunct}
{\mcitedefaultendpunct}{\mcitedefaultseppunct}\relax
\EndOfBibitem
\bibitem[Alvarez(2013)]{alvarez}
Alvarez,~S. A Cartography of the Van der {Waals} Territories. \emph{Dalton Trans.} \textbf{2013}, \emph{42}, 8617--8636\relax
\mciteBstWouldAddEndPuncttrue
\mciteSetBstMidEndSepPunct{\mcitedefaultmidpunct}
{\mcitedefaultendpunct}{\mcitedefaultseppunct}\relax
\EndOfBibitem
\bibitem[Lee \latin{et~al.}(2010)Lee, Murray, Kong, Lundqvist, and Langreth]{lee2010}
Lee,~K.; Murray,~{\'E}.~D.; Kong,~L.; Lundqvist,~B.~I.; Langreth,~D.~C. {Higher-Accuracy Van der Waals Density Functional}. \emph{Phys. Rev. B} \textbf{2010}, \emph{82}, 081101\relax
\mciteBstWouldAddEndPuncttrue
\mciteSetBstMidEndSepPunct{\mcitedefaultmidpunct}
{\mcitedefaultendpunct}{\mcitedefaultseppunct}\relax
\EndOfBibitem
\bibitem[Cooper(2010)]{cooper2010}
Cooper,~V.~R. {Van der Waals Density Functional: an Appropriate Exchange Functional}. \emph{Phys. Rev. B} \textbf{2010}, \emph{81}, 161104\relax
\mciteBstWouldAddEndPuncttrue
\mciteSetBstMidEndSepPunct{\mcitedefaultmidpunct}
{\mcitedefaultendpunct}{\mcitedefaultseppunct}\relax
\EndOfBibitem
\bibitem[Hamada and Otani(2010)Hamada, and Otani]{hamada2010}
Hamada,~I.; Otani,~M. {Comparative van der Waals Density-Functional Study of Graphene on Metal Surfaces}. \emph{Phys. Rev. B} \textbf{2010}, \emph{82}, 153412\relax
\mciteBstWouldAddEndPuncttrue
\mciteSetBstMidEndSepPunct{\mcitedefaultmidpunct}
{\mcitedefaultendpunct}{\mcitedefaultseppunct}\relax
\EndOfBibitem
\bibitem[Perdew \latin{et~al.}(1996)Perdew, Burke, and Ernzerhof]{PBE}
Perdew,~J.~P.; Burke,~K.; Ernzerhof,~M. Generalized Gradient Approximation Made Simple. \emph{Phys. Rev. Lett.} \textbf{1996}, \emph{77}, 3865--3868\relax
\mciteBstWouldAddEndPuncttrue
\mciteSetBstMidEndSepPunct{\mcitedefaultmidpunct}
{\mcitedefaultendpunct}{\mcitedefaultseppunct}\relax
\EndOfBibitem
\bibitem[Kozinsky and Marzari(2006)Kozinsky, and Marzari]{kozinsky2006static}
Kozinsky,~B.; Marzari,~N. Static Dielectric Properties of Carbon Nanotubes from First Principles. \emph{Phys. Rev. Lett.} \textbf{2006}, \emph{96}, 166801\relax
\mciteBstWouldAddEndPuncttrue
\mciteSetBstMidEndSepPunct{\mcitedefaultmidpunct}
{\mcitedefaultendpunct}{\mcitedefaultseppunct}\relax
\EndOfBibitem
\bibitem[Peierls(1930)]{peierls1930theorie}
Peierls,~R. Zur Theorie der Elektrischen und Thermischen Leitf{\"a}higkeit von Metallen. \emph{Ann. Phys.} \textbf{1930}, \emph{396}, 121--148\relax
\mciteBstWouldAddEndPuncttrue
\mciteSetBstMidEndSepPunct{\mcitedefaultmidpunct}
{\mcitedefaultendpunct}{\mcitedefaultseppunct}\relax
\EndOfBibitem
\bibitem[Pouget(2016)]{pouget2016peierls}
Pouget,~J.-P. The Peierls Instability and Charge Density Wave in One-Dimensional Electronic Conductors. \emph{C. R. Phys.} \textbf{2016}, \emph{17}, 332--356\relax
\mciteBstWouldAddEndPuncttrue
\mciteSetBstMidEndSepPunct{\mcitedefaultmidpunct}
{\mcitedefaultendpunct}{\mcitedefaultseppunct}\relax
\EndOfBibitem
\bibitem[Khomskii(2010)]{khomskii2010basic}
Khomskii,~D.~I. \emph{Basic Aspects of the Quantum Theory of Solids: Order and Elementary Excitations}; Cambridge University Press, 2010\relax
\mciteBstWouldAddEndPuncttrue
\mciteSetBstMidEndSepPunct{\mcitedefaultmidpunct}
{\mcitedefaultendpunct}{\mcitedefaultseppunct}\relax
\EndOfBibitem
\bibitem[Gr\"uner(1988)]{gruner1988thedyn}
Gr\"uner,~G. The Dynamics of Charge-Density Waves. \emph{Rev. Mod. Phys.} \textbf{1988}, \emph{60}, 1129--1181\relax
\mciteBstWouldAddEndPuncttrue
\mciteSetBstMidEndSepPunct{\mcitedefaultmidpunct}
{\mcitedefaultendpunct}{\mcitedefaultseppunct}\relax
\EndOfBibitem
\bibitem[Fr{\"o}hlich(1954)]{frohlich1954theory}
Fr{\"o}hlich,~H. On the theory of superconductivity: the one-dimensional case. \emph{Proceedings of the Royal Society of London. Series A. Mathematical and Physical Sciences} \textbf{1954}, \emph{223}, 296--305\relax
\mciteBstWouldAddEndPuncttrue
\mciteSetBstMidEndSepPunct{\mcitedefaultmidpunct}
{\mcitedefaultendpunct}{\mcitedefaultseppunct}\relax
\EndOfBibitem
\bibitem[Kohn(1959)]{kohn1959image}
Kohn,~W. Image of the Fermi Surface in the Vibration Spectrum of a Metal. \emph{Phys. Rev. Lett.} \textbf{1959}, \emph{2}, 393\relax
\mciteBstWouldAddEndPuncttrue
\mciteSetBstMidEndSepPunct{\mcitedefaultmidpunct}
{\mcitedefaultendpunct}{\mcitedefaultseppunct}\relax
\EndOfBibitem
\bibitem[Zhu \latin{et~al.}(2015)Zhu, Cao, Zhang, Plummer, and Guo]{zhu2015classification}
Zhu,~X.; Cao,~Y.; Zhang,~J.; Plummer,~E.; Guo,~J. Classification of Charge Density Waves Based on Their Nature. \emph{Proc. Natl. Acad. Sci.} \textbf{2015}, \emph{112}, 2367--2371\relax
\mciteBstWouldAddEndPuncttrue
\mciteSetBstMidEndSepPunct{\mcitedefaultmidpunct}
{\mcitedefaultendpunct}{\mcitedefaultseppunct}\relax
\EndOfBibitem
\bibitem[Zhu \latin{et~al.}(2017)Zhu, Guo, Zhang, and Plummer]{zhu2017misconceptions}
Zhu,~X.; Guo,~J.; Zhang,~J.; Plummer,~E. Misconceptions Associated with the Origin of Charge Density Waves. \emph{Adv. Phys.: X} \textbf{2017}, \emph{2}, 622--640\relax
\mciteBstWouldAddEndPuncttrue
\mciteSetBstMidEndSepPunct{\mcitedefaultmidpunct}
{\mcitedefaultendpunct}{\mcitedefaultseppunct}\relax
\EndOfBibitem
\bibitem[Toombs(1978)]{toombs1978quasi}
Toombs,~G.~A. Quasi-One-Dimensional Conductors. \emph{Phys. Rep.} \textbf{1978}, \emph{40}, 181--240\relax
\mciteBstWouldAddEndPuncttrue
\mciteSetBstMidEndSepPunct{\mcitedefaultmidpunct}
{\mcitedefaultendpunct}{\mcitedefaultseppunct}\relax
\EndOfBibitem
\bibitem[Johannes \latin{et~al.}(2006)Johannes, Mazin, and Howells]{johannes2006fermi}
Johannes,~M.; Mazin,~I.; Howells,~C. Fermi-Surface Nesting and the Origin of the Charge-Density Wave in NbSe2. \emph{Phys. Rev. B} \textbf{2006}, \emph{73}, 205102\relax
\mciteBstWouldAddEndPuncttrue
\mciteSetBstMidEndSepPunct{\mcitedefaultmidpunct}
{\mcitedefaultendpunct}{\mcitedefaultseppunct}\relax
\EndOfBibitem
\bibitem[Derriche \latin{et~al.}(2022)Derriche, Elfimov, and Sawatzky]{Derriche2022Suppression}
Derriche,~N.; Elfimov,~I.; Sawatzky,~G. Suppression of Peierls-Like Nesting-Based Instabilities in Solids. \emph{Phys. Rev. B} \textbf{2022}, \emph{106}, 064102\relax
\mciteBstWouldAddEndPuncttrue
\mciteSetBstMidEndSepPunct{\mcitedefaultmidpunct}
{\mcitedefaultendpunct}{\mcitedefaultseppunct}\relax
\EndOfBibitem
\bibitem[Johannes and Mazin(2008)Johannes, and Mazin]{johannes2008fermi}
Johannes,~M.; Mazin,~I. Fermi Surface Nesting and the Origin of Charge Density Waves in Metals. \emph{Phys. Rev. B} \textbf{2008}, \emph{77}, 165135\relax
\mciteBstWouldAddEndPuncttrue
\mciteSetBstMidEndSepPunct{\mcitedefaultmidpunct}
{\mcitedefaultendpunct}{\mcitedefaultseppunct}\relax
\EndOfBibitem
\bibitem[Calandra \latin{et~al.}(2009)Calandra, Mazin, and Mauri]{calandra2009effect}
Calandra,~M.; Mazin,~I.; Mauri,~F. Effect of Dimensionality on the Charge-Density Wave in Few-Layer 2H-NbSe2. \emph{Phys. Rev. B} \textbf{2009}, \emph{80}, 241108\relax
\mciteBstWouldAddEndPuncttrue
\mciteSetBstMidEndSepPunct{\mcitedefaultmidpunct}
{\mcitedefaultendpunct}{\mcitedefaultseppunct}\relax
\EndOfBibitem
\bibitem[Diego \latin{et~al.}(2021)Diego, Said, Mahatha, Bianco, Monacelli, Calandra, Mauri, Rossnagel, Errea, and Blanco-Canosa]{diego2021van}
Diego,~J.; Said,~A.; Mahatha,~S.~K.; Bianco,~R.; Monacelli,~L.; Calandra,~M.; Mauri,~F.; Rossnagel,~K.; Errea,~I.; Blanco-Canosa,~S. Van der Waals Driven Anharmonic Melting of the 3D Charge Density Wave in VSe2. \emph{Nat. Commun.} \textbf{2021}, \emph{12}, 1--7\relax
\mciteBstWouldAddEndPuncttrue
\mciteSetBstMidEndSepPunct{\mcitedefaultmidpunct}
{\mcitedefaultendpunct}{\mcitedefaultseppunct}\relax
\EndOfBibitem
\bibitem[Baroni \latin{et~al.}(2001)Baroni, De~Gironcoli, Dal~Corso, and Giannozzi]{baroni2001phonons}
Baroni,~S.; De~Gironcoli,~S.; Dal~Corso,~A.; Giannozzi,~P. Phonons and Related Crystal Properties from Density-Functional Perturbation Theory. \emph{Rev. Mod. Phys.} \textbf{2001}, \emph{73}, 515\relax
\mciteBstWouldAddEndPuncttrue
\mciteSetBstMidEndSepPunct{\mcitedefaultmidpunct}
{\mcitedefaultendpunct}{\mcitedefaultseppunct}\relax
\EndOfBibitem
\bibitem[Bockrath \latin{et~al.}(1999)Bockrath, Cobden, Lu, Rinzler, Smalley, Balents, and McEuen]{bockrath1999luttinger}
Bockrath,~M.; Cobden,~D.~H.; Lu,~J.; Rinzler,~A.~G.; Smalley,~R.~E.; Balents,~L.; McEuen,~P.~L. Luttinger-Liquid Behaviour in Carbon Nanotubes. \emph{Nature} \textbf{1999}, \emph{397}, 598--601\relax
\mciteBstWouldAddEndPuncttrue
\mciteSetBstMidEndSepPunct{\mcitedefaultmidpunct}
{\mcitedefaultendpunct}{\mcitedefaultseppunct}\relax
\EndOfBibitem
\bibitem[Chang(2003)]{Chang2003chiral}
Chang,~A. Chiral Luttinger Liquids at the Fractional Quantum Hall Edge. \emph{Rev. Mod. Phys.} \textbf{2003}, \emph{75}, 1449\relax
\mciteBstWouldAddEndPuncttrue
\mciteSetBstMidEndSepPunct{\mcitedefaultmidpunct}
{\mcitedefaultendpunct}{\mcitedefaultseppunct}\relax
\EndOfBibitem
\bibitem[Blumenstein \latin{et~al.}(2011)Blumenstein, Sch{\"a}fer, Mietke, Meyer, Dollinger, Lochner, Cui, Patthey, Matzdorf, and Claessen]{blumenstein2011atomically}
Blumenstein,~C.; Sch{\"a}fer,~J.; Mietke,~S.; Meyer,~S.; Dollinger,~A.; Lochner,~M.; Cui,~X.; Patthey,~L.; Matzdorf,~R.; Claessen,~R. Atomically Controlled Quantum Chains Hosting a Tomonaga--Luttinger Liquid. \emph{Nat. Phys.} \textbf{2011}, \emph{7}, 776--780\relax
\mciteBstWouldAddEndPuncttrue
\mciteSetBstMidEndSepPunct{\mcitedefaultmidpunct}
{\mcitedefaultendpunct}{\mcitedefaultseppunct}\relax
\EndOfBibitem
\bibitem[Bianco \latin{et~al.}(2019)Bianco, Errea, Monacelli, Calandra, and Mauri]{bianco2019quantum}
Bianco,~R.; Errea,~I.; Monacelli,~L.; Calandra,~M.; Mauri,~F. Quantum Enhancement of Charge Density Wave in NbS2 in the Two-Dimensional Limit. \emph{Nano Lett.} \textbf{2019}, \emph{19}, 3098--3103\relax
\mciteBstWouldAddEndPuncttrue
\mciteSetBstMidEndSepPunct{\mcitedefaultmidpunct}
{\mcitedefaultendpunct}{\mcitedefaultseppunct}\relax
\EndOfBibitem
\bibitem[Bianco \latin{et~al.}(2020)Bianco, Monacelli, Calandra, Mauri, and Errea]{bianco2020weak}
Bianco,~R.; Monacelli,~L.; Calandra,~M.; Mauri,~F.; Errea,~I. Weak Dimensionality Dependence and Dominant Role of Ionic Fluctuations in the Charge-Density-Wave Transition of NbSe2. \emph{Phys. Rev. Lett.} \textbf{2020}, \emph{125}, 106101\relax
\mciteBstWouldAddEndPuncttrue
\mciteSetBstMidEndSepPunct{\mcitedefaultmidpunct}
{\mcitedefaultendpunct}{\mcitedefaultseppunct}\relax
\EndOfBibitem
\bibitem[Cococcioni \latin{et~al.}(2005)Cococcioni, Mauri, Ceder, and Marzari]{cococcioni2005electronic}
Cococcioni,~M.; Mauri,~F.; Ceder,~G.; Marzari,~N. Electronic-Enthalpy Functional for Finite Systems Under Pressure. \emph{Phys. Rev. Lett.} \textbf{2005}, \emph{94}, 145501\relax
\mciteBstWouldAddEndPuncttrue
\mciteSetBstMidEndSepPunct{\mcitedefaultmidpunct}
{\mcitedefaultendpunct}{\mcitedefaultseppunct}\relax
\EndOfBibitem
\bibitem[Liu \latin{et~al.}(2016)Liu, Debnath, Pope, Salguero, Lake, and Balandin]{liu2016charge}
Liu,~G.; Debnath,~B.; Pope,~T.~R.; Salguero,~T.~T.; Lake,~R.~K.; Balandin,~A.~A. A Charge-Density-Wave Oscillator Based on an Itegrated Tantalum Disulfide--Boron Nitride--Graphene Device Operating at Room Temperature. \emph{Nat. Nanotechnol.} \textbf{2016}, \emph{11}, 845--850\relax
\mciteBstWouldAddEndPuncttrue
\mciteSetBstMidEndSepPunct{\mcitedefaultmidpunct}
{\mcitedefaultendpunct}{\mcitedefaultseppunct}\relax
\EndOfBibitem
\bibitem[Zhu \latin{et~al.}(2018)Zhu, Chen, Liu, Zheng, Li, Chaturvedi, Zhou, Fu, He, Zeng, and et~al.]{zhu2018light}
Zhu,~C.; Chen,~Y.; Liu,~F.; Zheng,~S.; Li,~X.; Chaturvedi,~A.; Zhou,~J.; Fu,~Q.; He,~Y.; Zeng,~Q.; et~al. Light-Tunable 1T-TaS2 Charge-Density-Wave Oscillators. \emph{ACS Nano} \textbf{2018}, \emph{12}, 11203--11210\relax
\mciteBstWouldAddEndPuncttrue
\mciteSetBstMidEndSepPunct{\mcitedefaultmidpunct}
{\mcitedefaultendpunct}{\mcitedefaultseppunct}\relax
\EndOfBibitem
\bibitem[Geremew \latin{et~al.}(2019)Geremew, Rumyantsev, Kargar, Debnath, Nosek, Bloodgood, Bockrath, Salguero, Lake, and Balandin]{geremew2019bias}
Geremew,~A.~K.; Rumyantsev,~S.; Kargar,~F.; Debnath,~B.; Nosek,~A.; Bloodgood,~M.~A.; Bockrath,~M.; Salguero,~T.~T.; Lake,~R.~K.; Balandin,~A.~A. Bias-Voltage Driven Switching of the Charge-Density-Wave and Normal Metallic Phases in 1T-TaS2 Thin-Film Devices. \emph{ACS Nano} \textbf{2019}, \emph{13}, 7231--7240\relax
\mciteBstWouldAddEndPuncttrue
\mciteSetBstMidEndSepPunct{\mcitedefaultmidpunct}
{\mcitedefaultendpunct}{\mcitedefaultseppunct}\relax
\EndOfBibitem
\bibitem[Mohammadzadeh \latin{et~al.}(2021)Mohammadzadeh, Baraghani, Yin, Kargar, Bird, and Balandin]{mohammadzadeh2021evidence}
Mohammadzadeh,~A.; Baraghani,~S.; Yin,~S.; Kargar,~F.; Bird,~J.~P.; Balandin,~A.~A. Evidence for a Thermally Driven Charge-Density-Wave Transition in 1T-TaS2 Thin-Film Devices: Prospects for GHz Switching Speed. \emph{Appl. Phys. Lett.} \textbf{2021}, \emph{118}, 093102\relax
\mciteBstWouldAddEndPuncttrue
\mciteSetBstMidEndSepPunct{\mcitedefaultmidpunct}
{\mcitedefaultendpunct}{\mcitedefaultseppunct}\relax
\EndOfBibitem
\bibitem[Gao \latin{et~al.}(2022)Gao, Guo, Wang, Nielsen, and Baughman]{gao2022strongest}
Gao,~E.; Guo,~Y.; Wang,~Z.; Nielsen,~S.~O.; Baughman,~R.~H. The Strongest and Toughest Predicted Materials: Linear Atomic Chains Without a Peierls Instability. \emph{Matter} \textbf{2022}, \emph{5}, 1192--1203\relax
\mciteBstWouldAddEndPuncttrue
\mciteSetBstMidEndSepPunct{\mcitedefaultmidpunct}
{\mcitedefaultendpunct}{\mcitedefaultseppunct}\relax
\EndOfBibitem
\bibitem[Treacy \latin{et~al.}(1996)Treacy, Ebbesen, and Gibson]{treacy1996exceptionally}
Treacy,~M.~J.; Ebbesen,~T.~W.; Gibson,~J.~M. Exceptionally High Young's Modulus Observed for Individual Carbon Nanotubes. \emph{Nature} \textbf{1996}, \emph{381}, 678--680\relax
\mciteBstWouldAddEndPuncttrue
\mciteSetBstMidEndSepPunct{\mcitedefaultmidpunct}
{\mcitedefaultendpunct}{\mcitedefaultseppunct}\relax
\EndOfBibitem
\bibitem[Leu \latin{et~al.}(2008)Leu, Svizhenko, and Cho]{leu2008ab}
Leu,~P.~W.; Svizhenko,~A.; Cho,~K. Ab Initio Calculations of the Mechanical and Electronic Properties of Strained Si Nanowires. \emph{Phys. Rev. B} \textbf{2008}, \emph{77}, 235305\relax
\mciteBstWouldAddEndPuncttrue
\mciteSetBstMidEndSepPunct{\mcitedefaultmidpunct}
{\mcitedefaultendpunct}{\mcitedefaultseppunct}\relax
\EndOfBibitem
\bibitem[Tritt \latin{et~al.}(1994)Tritt, Jacobsen, Ehrlich, and Gillespie]{tritt1994measure}
Tritt,~T.~M.; Jacobsen,~R.~L.; Ehrlich,~A.~C.; Gillespie,~D.~J. Measure of the Elastic and Transport Properties of TaSe3 Through the Stress-Induced Phase Transition. \emph{Phys. B} \textbf{1994}, \emph{194}, 1303--1304\relax
\mciteBstWouldAddEndPuncttrue
\mciteSetBstMidEndSepPunct{\mcitedefaultmidpunct}
{\mcitedefaultendpunct}{\mcitedefaultseppunct}\relax
\EndOfBibitem
\bibitem[Liu \latin{et~al.}(2015)Liu, Xu, Chen, and Shen]{liu2015flexible}
Liu,~Z.; Xu,~J.; Chen,~D.; Shen,~G. Flexible Electronics Based on Inorganic Nanowires. \emph{Chem. Soc. Rev.} \textbf{2015}, \emph{44}, 161--192\relax
\mciteBstWouldAddEndPuncttrue
\mciteSetBstMidEndSepPunct{\mcitedefaultmidpunct}
{\mcitedefaultendpunct}{\mcitedefaultseppunct}\relax
\EndOfBibitem
\bibitem[Dag \latin{et~al.}(2005)Dag, Tongay, Yildirim, Durgun, Senger, Fong, and Ciraci]{dag2005half}
Dag,~S.; Tongay,~S.; Yildirim,~T.; Durgun,~E.; Senger,~R.; Fong,~C.; Ciraci,~S. Half-Metallic Properties of Atomic Chains of Carbon--Transition-Metal Compounds. \emph{Phys. Rev. B} \textbf{2005}, \emph{72}, 155444\relax
\mciteBstWouldAddEndPuncttrue
\mciteSetBstMidEndSepPunct{\mcitedefaultmidpunct}
{\mcitedefaultendpunct}{\mcitedefaultseppunct}\relax
\EndOfBibitem
\bibitem[Sun \latin{et~al.}(2016)Sun, Cai, Wang, Widmer, Ju, Zhu, Li, He, Ruffieux, Fasel, and et~al.]{sun2016bottom}
Sun,~Q.; Cai,~L.; Wang,~S.; Widmer,~R.; Ju,~H.; Zhu,~J.; Li,~L.; He,~Y.; Ruffieux,~P.; Fasel,~R.; et~al. Bottom-Up Synthesis of Metalated Carbyne. \emph{J. Am. Chem. Soc.} \textbf{2016}, \emph{138}, 1106--1109\relax
\mciteBstWouldAddEndPuncttrue
\mciteSetBstMidEndSepPunct{\mcitedefaultmidpunct}
{\mcitedefaultendpunct}{\mcitedefaultseppunct}\relax
\EndOfBibitem
\bibitem[Tu \latin{et~al.}(2016)Tu, Wang, Shen, Wang, Sanvito, and Hou]{tu2016cu}
Tu,~X.; Wang,~H.; Shen,~Z.; Wang,~Y.; Sanvito,~S.; Hou,~S. Cu-Metalated Carbyne Acting As a Promising Molecular Wire. \emph{J. Chem. Phys.} \textbf{2016}, \emph{145}, 244702\relax
\mciteBstWouldAddEndPuncttrue
\mciteSetBstMidEndSepPunct{\mcitedefaultmidpunct}
{\mcitedefaultendpunct}{\mcitedefaultseppunct}\relax
\EndOfBibitem
\bibitem[Ruschewitz(2003)]{ruschewitz2003binary}
Ruschewitz,~U. Binary and Ternary Carbides of Alkali and Alkaline-Earth Metals. \emph{Coord. Chem. Rev.} \textbf{2003}, \emph{244}, 115--136\relax
\mciteBstWouldAddEndPuncttrue
\mciteSetBstMidEndSepPunct{\mcitedefaultmidpunct}
{\mcitedefaultendpunct}{\mcitedefaultseppunct}\relax
\EndOfBibitem
\bibitem[Cremer \latin{et~al.}(2002)Cremer, Kockelmann, Bertmer, and Ruschewitz]{cremer2002alkali}
Cremer,~U.; Kockelmann,~W.; Bertmer,~M.; Ruschewitz,~U. Alkali Metal Copper Acetylides ACuC2 (A= Na--Cs): Synthesis, Crystal Structures and Spectroscopic Properties. \emph{Solid State Sci.} \textbf{2002}, \emph{4}, 247--253\relax
\mciteBstWouldAddEndPuncttrue
\mciteSetBstMidEndSepPunct{\mcitedefaultmidpunct}
{\mcitedefaultendpunct}{\mcitedefaultseppunct}\relax
\EndOfBibitem
\bibitem[Jin and Liu(2020)Jin, and Liu]{jin20201d}
Jin,~K.-H.; Liu,~F. 1D Topological Phases in Transition--Metal Monochalcogenide Nanowires. \emph{Nanoscale} \textbf{2020}, \emph{12}, 14661--14667\relax
\mciteBstWouldAddEndPuncttrue
\mciteSetBstMidEndSepPunct{\mcitedefaultmidpunct}
{\mcitedefaultendpunct}{\mcitedefaultseppunct}\relax
\EndOfBibitem
\bibitem[Liu \latin{et~al.}(2022)Liu, Yin, Singh, and Liu]{liu2022ta}
Liu,~S.; Yin,~H.; Singh,~D.~J.; Liu,~P.-F. Ta4SiTe4: A Possible One-Dimensional Topological Insulator. \emph{Phys. Rev. B} \textbf{2022}, \emph{105}, 195419\relax
\mciteBstWouldAddEndPuncttrue
\mciteSetBstMidEndSepPunct{\mcitedefaultmidpunct}
{\mcitedefaultendpunct}{\mcitedefaultseppunct}\relax
\EndOfBibitem
\bibitem[Varsano \latin{et~al.}(2020)Varsano, Palummo, Molinari, and Rontani]{varsano2020monolayer}
Varsano,~D.; Palummo,~M.; Molinari,~E.; Rontani,~M. A Monolayer Transition-Metal Dichalcogenide as a Topological Excitonic Insulator. \emph{Nat. Nanotechnol.} \textbf{2020}, \emph{15}, 367--372\relax
\mciteBstWouldAddEndPuncttrue
\mciteSetBstMidEndSepPunct{\mcitedefaultmidpunct}
{\mcitedefaultendpunct}{\mcitedefaultseppunct}\relax
\EndOfBibitem
\bibitem[J{\'e}rome \latin{et~al.}(1967)J{\'e}rome, Rice, and Kohn]{jerome1967excitonic}
J{\'e}rome,~D.; Rice,~T.; Kohn,~W. Excitonic Insulator. \emph{Phys. Rev.} \textbf{1967}, \emph{158}, 462\relax
\mciteBstWouldAddEndPuncttrue
\mciteSetBstMidEndSepPunct{\mcitedefaultmidpunct}
{\mcitedefaultendpunct}{\mcitedefaultseppunct}\relax
\EndOfBibitem
\bibitem[Geim(2009)]{geim2009graphene}
Geim,~A.~K. Graphene: Status and Prospects. \emph{Science} \textbf{2009}, \emph{324}, 1530--1534\relax
\mciteBstWouldAddEndPuncttrue
\mciteSetBstMidEndSepPunct{\mcitedefaultmidpunct}
{\mcitedefaultendpunct}{\mcitedefaultseppunct}\relax
\EndOfBibitem
\bibitem[Yamamoto(1978)]{yamamoto1978superconducting}
Yamamoto,~M. Superconducting Properties of TaSe3. \emph{J. Phys. Soc. Jpn.} \textbf{1978}, \emph{45}, 431--438\relax
\mciteBstWouldAddEndPuncttrue
\mciteSetBstMidEndSepPunct{\mcitedefaultmidpunct}
{\mcitedefaultendpunct}{\mcitedefaultseppunct}\relax
\EndOfBibitem
\bibitem[Nagata \latin{et~al.}(1989)Nagata, Kutsuzawa, Ebisu, Yamamura, and Taniguchi]{nagata1989superconductivity}
Nagata,~S.; Kutsuzawa,~H.; Ebisu,~S.; Yamamura,~H.; Taniguchi,~S. Superconductivity in the Quasi-One-Dimensional Conductor TaSe3. \emph{J. Phys. Chem. Solids} \textbf{1989}, \emph{50}, 703--707\relax
\mciteBstWouldAddEndPuncttrue
\mciteSetBstMidEndSepPunct{\mcitedefaultmidpunct}
{\mcitedefaultendpunct}{\mcitedefaultseppunct}\relax
\EndOfBibitem
\bibitem[Giannozzi \latin{et~al.}(2009)Giannozzi, Baroni, Bonini, Calandra, Car, Cavazzoni, Ceresoli, Chiarotti, Cococcioni, Dabo, and et~al.]{giannozzi2009quantum}
Giannozzi,~P.; Baroni,~S.; Bonini,~N.; Calandra,~M.; Car,~R.; Cavazzoni,~C.; Ceresoli,~D.; Chiarotti,~G.~L.; Cococcioni,~M.; Dabo,~I.; et~al. QUANTUM ESPRESSO: a Modular and Open-Source Software Project for Quantum Simulations of Materials. \emph{J. Phys.: Condens.Matter} \textbf{2009}, \emph{21}, 395502\relax
\mciteBstWouldAddEndPuncttrue
\mciteSetBstMidEndSepPunct{\mcitedefaultmidpunct}
{\mcitedefaultendpunct}{\mcitedefaultseppunct}\relax
\EndOfBibitem
\bibitem[Prandini \latin{et~al.}(2018)Prandini, Marrazzo, Castelli, Mounet, and Marzari]{prandini2018precision}
Prandini,~G.; Marrazzo,~A.; Castelli,~I.~E.; Mounet,~N.; Marzari,~N. Precision and Efficiency in Solid-State Pseudopotential Calculations. \emph{npj Comput. Mater.} \textbf{2018}, \emph{4}, 1--13\relax
\mciteBstWouldAddEndPuncttrue
\mciteSetBstMidEndSepPunct{\mcitedefaultmidpunct}
{\mcitedefaultendpunct}{\mcitedefaultseppunct}\relax
\EndOfBibitem
\bibitem[Dal~Corso(2014)]{dal2014pseudopotentials}
Dal~Corso,~A. Pseudopotentials Periodic Table: From H to Pu. \emph{Comput. Mater. Sci.} \textbf{2014}, \emph{95}, 337--350\relax
\mciteBstWouldAddEndPuncttrue
\mciteSetBstMidEndSepPunct{\mcitedefaultmidpunct}
{\mcitedefaultendpunct}{\mcitedefaultseppunct}\relax
\EndOfBibitem
\bibitem[Marzari \latin{et~al.}(1999)Marzari, Vanderbilt, De~Vita, and Payne]{marzari1999thermal}
Marzari,~N.; Vanderbilt,~D.; De~Vita,~A.; Payne,~M. Thermal Contraction and Disordering of the Al (110) Surface. \emph{Phys. Rev. Lett.} \textbf{1999}, \emph{82}, 3296\relax
\mciteBstWouldAddEndPuncttrue
\mciteSetBstMidEndSepPunct{\mcitedefaultmidpunct}
{\mcitedefaultendpunct}{\mcitedefaultseppunct}\relax
\EndOfBibitem
\bibitem[Lin \latin{et~al.}(2022)Lin, Ponc{\'e}, and Marzari]{lin2022general}
Lin,~C.; Ponc{\'e},~S.; Marzari,~N. General Invariance and Equilibrium Conditions for Lattice Dynamics in 1D, 2D, and 3D Materials. \emph{npj Comput. Mater.} \textbf{2022}, \emph{8}, 236\relax
\mciteBstWouldAddEndPuncttrue
\mciteSetBstMidEndSepPunct{\mcitedefaultmidpunct}
{\mcitedefaultendpunct}{\mcitedefaultseppunct}\relax
\EndOfBibitem
\bibitem[Andreussi \latin{et~al.}(2012)Andreussi, Dabo, and Marzari]{andreussi2012revised}
Andreussi,~O.; Dabo,~I.; Marzari,~N. Revised Self-Consistent Continuum Solvation in Electronic-Structure Calculations. \emph{J. Chem. Phys.} \textbf{2012}, \emph{136}, 064102\relax
\mciteBstWouldAddEndPuncttrue
\mciteSetBstMidEndSepPunct{\mcitedefaultmidpunct}
{\mcitedefaultendpunct}{\mcitedefaultseppunct}\relax
\EndOfBibitem
\bibitem[Zhang \latin{et~al.}(2011)Zhang, Su, Wang, Kong, Chen, and Zhang]{zhang2011one}
Zhang,~Y.; Su,~Y.; Wang,~L.; Kong,~E. S.-W.; Chen,~X.; Zhang,~Y. A One-Dimensional Extremely Covalent Material: Monatomic Carbon Linear Chain. \emph{Nanoscale Res. Lett.} \textbf{2011}, \emph{6}, 1--4\relax
\mciteBstWouldAddEndPuncttrue
\mciteSetBstMidEndSepPunct{\mcitedefaultmidpunct}
{\mcitedefaultendpunct}{\mcitedefaultseppunct}\relax
\EndOfBibitem
\bibitem[Min \latin{et~al.}(2022)Min, Zhuang, and Yao]{min2022half}
Min,~Y.; Zhuang,~G.; Yao,~K. Half-Metallicity in Cu-Metalated Carbyne Predicted by First-Principles Calculations. \emph{Phys. Lett. A} \textbf{2022}, \emph{449}, 128357\relax
\mciteBstWouldAddEndPuncttrue
\mciteSetBstMidEndSepPunct{\mcitedefaultmidpunct}
{\mcitedefaultendpunct}{\mcitedefaultseppunct}\relax
\EndOfBibitem
\bibitem[Sabatini \latin{et~al.}(2013)Sabatini, Gorni, and De~Gironcoli]{sabatini2013nonlocal}
Sabatini,~R.; Gorni,~T.; De~Gironcoli,~S. Nonlocal Van der Waals Density Functional Made Simple and Efficient. \emph{Phys. Rev. B} \textbf{2013}, \emph{87}, 041108\relax
\mciteBstWouldAddEndPuncttrue
\mciteSetBstMidEndSepPunct{\mcitedefaultmidpunct}
{\mcitedefaultendpunct}{\mcitedefaultseppunct}\relax
\EndOfBibitem
\bibitem[Heyd \latin{et~al.}(2003)Heyd, Scuseria, and Ernzerhof]{heyd2003hybrid}
Heyd,~J.; Scuseria,~G.~E.; Ernzerhof,~M. Hybrid functionals based on a screened Coulomb Potential. \emph{J. Chem. Phys.} \textbf{2003}, \emph{118}, 8207--8215\relax
\mciteBstWouldAddEndPuncttrue
\mciteSetBstMidEndSepPunct{\mcitedefaultmidpunct}
{\mcitedefaultendpunct}{\mcitedefaultseppunct}\relax
\EndOfBibitem
\bibitem[van Setten \latin{et~al.}(2018)van Setten, Giantomassi, Bousquet, Verstraete, Hamann, Gonze, and Rignanese]{van2018pseudodojo}
van Setten,~M.~J.; Giantomassi,~M.; Bousquet,~E.; Verstraete,~M.~J.; Hamann,~D.~R.; Gonze,~X.; Rignanese,~G.-M. The PseudoDojo: Training and Grading a 85 Element Optimized Norm-Conserving Pseudopotential Table. \emph{Comput. Phys. Commun.} \textbf{2018}, \emph{226}, 39--54\relax
\mciteBstWouldAddEndPuncttrue
\mciteSetBstMidEndSepPunct{\mcitedefaultmidpunct}
{\mcitedefaultendpunct}{\mcitedefaultseppunct}\relax
\EndOfBibitem
\bibitem[Talirz \latin{et~al.}(2020)Talirz, Kumbhar, Passaro, Yakutovich, Granata, Gargiulo, Borelli, Uhrin, Huber, Zoupanos, and et~al.]{talirz2020materials}
Talirz,~L.; Kumbhar,~S.; Passaro,~E.; Yakutovich,~A.~V.; Granata,~V.; Gargiulo,~F.; Borelli,~M.; Uhrin,~M.; Huber,~S.~P.; Zoupanos,~S.; et~al. Materials Cloud, a platform for open computational science. \emph{Scientific Data} \textbf{2020}, \emph{7}, 299\relax
\mciteBstWouldAddEndPuncttrue
\mciteSetBstMidEndSepPunct{\mcitedefaultmidpunct}
{\mcitedefaultendpunct}{\mcitedefaultseppunct}\relax
\EndOfBibitem
\bibitem[Cignarella \latin{et~al.}(2024)Cignarella, Campi, and Marzari]{cignarella2024cloud}
Cignarella,~C.; Campi,~D.; Marzari,~N. Searching for the thinnest metallic wire. \emph{Materials Cloud Archive} \textbf{2024}, 2024.32, DOI: 10.24435/materialscloud:xh-za.\relax
\mciteBstWouldAddEndPunctfalse
\mciteSetBstMidEndSepPunct{\mcitedefaultmidpunct}
{}{\mcitedefaultseppunct}\relax
\EndOfBibitem
\end{mcitethebibliography}
\end{document}